\documentclass[prb,twocolumn,10pt,aps,showpacs,amsmath,amsfonts,floatfix,letterpaper]{revtex4-1}

\usepackage{times}
\usepackage{txfonts}

\usepackage{amsmath}
\usepackage{amssymb}
\usepackage{hyperref}
\usepackage{multirow}

\newcommand{\ee}{\mathrm{e}}

\newcommand{\dd}{\mathrm{d}}
\newcommand{\DD}{\mathrm{D}}

\newcommand{\del}{\boldsymbol{\nabla}}

\newcommand{\Bv}{\boldsymbol{B}}

\newcommand{\rv}{\boldsymbol{r}}
\newcommand{\Mv}{\boldsymbol{M}}

\newcommand{\scH}{\mathcal{H}}
\newcommand{\scZ}{\mathcal{Z}}
\newcommand{\scz}{\mathfrak{z}}

\newcommand{\Sv}{\boldsymbol{S}}
\newcommand{\nhv}{\hat{\boldsymbol{n}}}
\newcommand{\Rv}{\boldsymbol{R}}
\newcommand{\Qv}{\boldsymbol{Q}}
\newcommand{\Pv}{\boldsymbol{P}}
\newcommand{\etahv}{\hat{\boldsymbol{\eta}}}
\newcommand{\scS}{\mathcal{S}}
\newcommand{\zerov}{\boldsymbol{0}}

\newcommand{\Tch}{T\sub{c$>$}}
\newcommand{\Tcl}{T\sub{c$<$}}

\newcommand{\HTO}{$\text{Ho}_2\text{Ti}_2\text{O}_7$}
\newcommand{\DTO}{$\text{Dy}_2\text{Ti}_2\text{O}_7$}

\newcommand{\beq}[1]{\begin{equation}\label{#1}}
\newcommand{\eeq}{\end{equation}}
\newcommand{\refeq}[1]{Eq.~(\ref{#1})}

\newcommand{\refcite}[1]{Ref.~\onlinecite{#1}}
\newcommand{\refcites}[1]{Refs.~\onlinecite{#1}}
\newcommand{\reffig}[1]{Fig.~\ref{#1}}
\newcommand{\reffigand}[2]{Figs.~\ref{#1} and \ref{#2}}
\newcommand{\refsec}[1]{Section~\ref{#1}}
\newcommand{\refsecand}[2]{Sections~\ref{#1} and \ref{#2}}

\newcommand{\punc}[1]{\,{\text{#1}}}
\newcommand{\sub}[1]{_{\text{#1}}}

\usepackage{color}

\usepackage{graphicx}
\usepackage{ifpdf}
\ifpdf

\newcommand{\putinscaledfigure}[1]{\begin{center}\includegraphics[width=\columnwidth]{#1.pdf}\end{center}}
\newcommand{\putinscaledwidefigure}[1]{\begin{center}\includegraphics[width=0.8\textwidth]{#1.pdf}\end{center}}
\else

\newcommand{\putinscaledfigure}[1]{\begin{center}\includegraphics[width=\columnwidth]{#1.eps}\end{center}}
\newcommand{\putinscaledwidefigure}[1]{\begin{center}\includegraphics[width=0.8\textwidth]{#1.eps}\end{center}}
\fi

\begin{document}

\title{Ferromagnetic Coulomb phase in classical spin ice}

\author{Stephen Powell}
\affiliation{School of Physics and Astronomy, The University of Nottingham, Nottingham, NG7 2RD, United Kingdom}

\begin{abstract}

Spin ice is a frustrated magnetic system that at low temperatures exhibits a Coulomb phase, a classical spin liquid with topological order and deconfined excitations. This work establishes the presence of a Coulomb phase with coexisting ferromagnetic order in a microscopic model of classical spin ice subject to uniaxial lattice distortion. General theoretical arguments are presented for the presence of such a phase, and its existence is confirmed using Monte Carlo results. This example is used to illustrate generic properties of spin liquids with magnetic order, including deconfinement of monopoles, signatures in the neutron-scattering structure factor, and critical behavior at phase transitions. An analogous phase, a superfluid with spontaneously broken particle--hole symmetry, is demonstrated in a model of hard-core lattice bosons, related to spin ice through the quantum--classical correspondence.

\end{abstract}

\maketitle

\section{Introduction}

Spin liquids are phases where magnetic degrees of freedom exhibit strong local correlations, despite the persistence of large fluctuations,\cite{BalentsReview} of either quantum mechanical or thermal origin. They occur at low temperature in certain frustrated systems, where interactions are large compared to thermal fluctuations, but mutual competition between the interactions prevents formation of a rigidly ordered configuration. Spin liquids have been of theoretical interest for several decades,\cite{Anderson1973,Anderson1987} but evidence for their existence in physical systems,\cite{Lee2008,Han2012} and even microscopic models,\cite{MoessnerSondhi2001,Balents2002,SpinIceReview} is considerably more recent.

While spin liquids are often distinguished from conventional low-temperature phases, such as ferromagnets, by the fact that they lack magnetic order, their defining characteristics go beyond the mere absence of conventional order. A precise definition of a quantum spin liquid (QSL) can be phrased in terms of long-range entanglement,\cite{BalentsReview} while the Coulomb phase,\cite{HenleyReview} the classical spin liquid (CSL) that is of primary interest here, can be defined through deconfinement of fractionalized ``monopole'' excitations.\cite{SpinIceReview} Experimental evidence exists for a Coulomb phase in the spin-ice compounds, which can be treated as classical at relevant temperatures.\cite{Bramwell}

These definitions provide positive characterizations for QSL and CSL phases, and also make clear the possibility of a magnetically ordered spin liquid, in which spin-liquid phenomena coexist with conventional symmetry-breaking order. Some examples of such phases have been reported in the theoretical literature: Mean-field studies of quantum spin ice\cite{SavaryBalents} identified an ordered QSL, referred to as a ``Coulomb ferromagnet'', although quantum Monte Carlo simulations have not revealed such a phase.\cite{Benton} Recent work\cite{BrooksBartlett} has also demonstrated the possibility of antiferromagnetic order coexisting with a CSL.

The compatibility of magnetic order and spin-liquid phenomenology also allows for the existence of phase transitions between ordered and disordered spin liquids. One might anticipate novel critical behavior at such transitions, since it is known that transitions from spin liquids into conventional ordered phases can transcend the usual Landau paradigm.\cite{HenleyReview}

This work demonstrates that a ferromagnetic Coulomb phase can occur in a model of classical spin ice, and provides a detailed study of this phase and the associated transitions. Theoretical arguments, including mapping to a related quantum model, are used to show that such a phase exists and that it can be reached through a continuous transition from the paramagnetic Coulomb phase. We present Monte Carlo (MC) results that confirm both of these statements, and illustrate the generic properties of ordered spin liquids, including the structure factor for elastic neutron scattering.

We also consider the critical behavior at the ordering transition and predict that, despite the Ising nature of the order parameter and the presence of only short-range interactions in the microscopic model, the transition should belong in the mean-field universality class, as a result of coupling to the effective gauge-field fluctuations of the spin liquid. While the numerical results are consistent with this prediction, larger system sizes would be required for a definitive confirmation of the universality class. This phase transition provides another interesting example of the diversity of critical phenomena that exists in the neighborhood of spin-liquid phases.

\subsection*{Outline}

In \refsec{SecModelAndPhases}, the model of spin ice is introduced, and a choice of perturbations that lead to a ferromagnetic Coulomb phase is motivated. The basic structure of the phase diagram is then illustrated using MC results, showing the appearance of such a phase at intermediate temperatures for certain values of the parameters. In \refsec{SecIntermediatePhase}, the phase in question is studied in detail, to confirm that it has nonzero magnetization while simultaneously exhibiting the characteristic features of a Coulomb phase. \refsecand{SecHigherTemperatureTransition}{SecLowerTemperatureTransition} consider in turn the critical behavior at the higher- and lower-temperature transitions out of this ferromagnetic Coulomb phase.

We conclude in \refsec{SecDiscussion}, by summarizing the features that are expected to be generic to ordered spin liquids, both quantum and classical, and discuss briefly the effect of a nonzero density of magnetic monopoles. In the Appendix, the classical--quantum mapping developed in \refcite{SpinIce100} is applied to this system, and the resulting quantum model is related to a problem of hard-core quantum bosons studied by Rokhsar and Kotliar.\cite{RokhsarKotliar}

\section{Model and phase structure}
\label{SecModelAndPhases}

\subsection{Nearest-neighbor model of spin ice}
\label{SecModel}

The spin-ice materials\cite{Bramwell,SpinIceReview} \HTO\ and \DTO\ are well described by a model of classical spins $\Sv_i$ on the sites $i$ of a pyrochlore lattice, a network of corner-sharing tetrahedra. Each spin is subject to a strong easy-axis anisotropy constraining it to point parallel or antiparallel to the local $\langle111\rangle$ axis joining the centres of adjacent tetrahedra, $\Sv_i = \pm\nhv_i$. Including only nearest-neighbor interactions, the Hamiltonian can be written as
\beq{EqHnn}
\scH\sub{nn} = -\sum_{\langle i j \rangle} J_{ij} \Sv_i \cdot \Sv_j\punc{,}
\eeq
where $J_{ij}$ is a ferromagnetic coupling between nearest-neighbor sites $\langle i j \rangle$ of the lattice.

In the unperturbed model, the interaction is uniform, $J_{ij} = J > 0$, and favors those states where, of the four spins on each tetrahedron, two point in and two point out. The latter condition is referred to as the ``ice rule'' and selects a set of states that is degenerate in the nearest-neighbor model and whose number grows exponentially with the number of spins. While a more realistic microscopic model than $\scH\sub{nn}$ would also include dipolar interactions,\cite{Bramwell} their effect is primarily to renormalize $J$, with only a small splitting of the ice-rule states.\cite{Isakov}

We will mostly concentrate on the limit where the ice rule is enforced as a constraint, represented by \refeq{EqHnn} with temperature $T \ll J$. Assuming ergodicity within the ice-rule manifold, the system in this limit exhibits a Coulomb phase, in which the spins are disordered but highly correlated. Replacing the spins $\Sv_i$ by a continuous vector field $\Bv(\rv)$ and the ice rule by $\del \cdot \Bv = 0$ leads to an effective coarse-grained description for this phase.\cite{HenleyReview} A quadratic action for the ``magnetic field'' $\Bv$ correctly describes the long-wavelength neutron scattering at low temperature in spin ice,\cite{Fennell} and predicts that monopoles in $\Bv$, corresponding to single tetrahedra where the ice rule is broken, are deconfined.\cite{SpinIceReview} Much of the physics is in fact qualitatively unaltered by a small density of such defects (see \refsec{SecDiscussion}), and their effects on the critical properties can be understood by treating monopole fugacity as a relevant perturbation (in the renormalization-group sense).\cite{MonopoleScalingPRL,MonopoleScalingPRB}

An important property of the ice-rule states for present purposes is that they obey a topological constraint on the magnetization:\cite{SpinIceReview,HenleyReview} Starting from any ice-rule state and flipping a spin $\Sv_i$ breaks the ice rule on the two tetrahedra shared by site $i$. The only updates that connect configurations within the ice-rule manifold are those that involve flipping a set of spins aligned head-to-tail along a closed loop. Any such update for a contractible loop preserves the magnetization density,
\beq{EqDefineMagnetization}
\Mv = \frac{1}{N\sub{s}}\sum_i \Sv_i\punc{,}
\eeq
where $N\sub{s} = \sum_i 1$ is the number of spins. Changing the magnetization while remaining within an ice-rule state in fact requires flipping spins along a loop that spans the system (assuming periodic boundary conditions). Sets of states with the same magnetization therefore constitute ``topological sectors'',\cite{SpinIceReview} disconnected by local updates. This topological conservation law is broken by a nonzero density of monopoles, but remains approximately valid, and conceptually useful, at low temperatures.

Nonzero magnetic susceptibility \(\chi\) requires that the system fluctuates between different sectors;\cite{Jaubert2013} in the thermodynamic limit, one can therefore distinguish ``incompressible'' phases with \(\chi = 0\) from those with \(\chi > 0\).

\subsection{Uniaxial distortion}
\label{SecUniaxialDistortion}

\begin{figure}
\putinscaledfigure{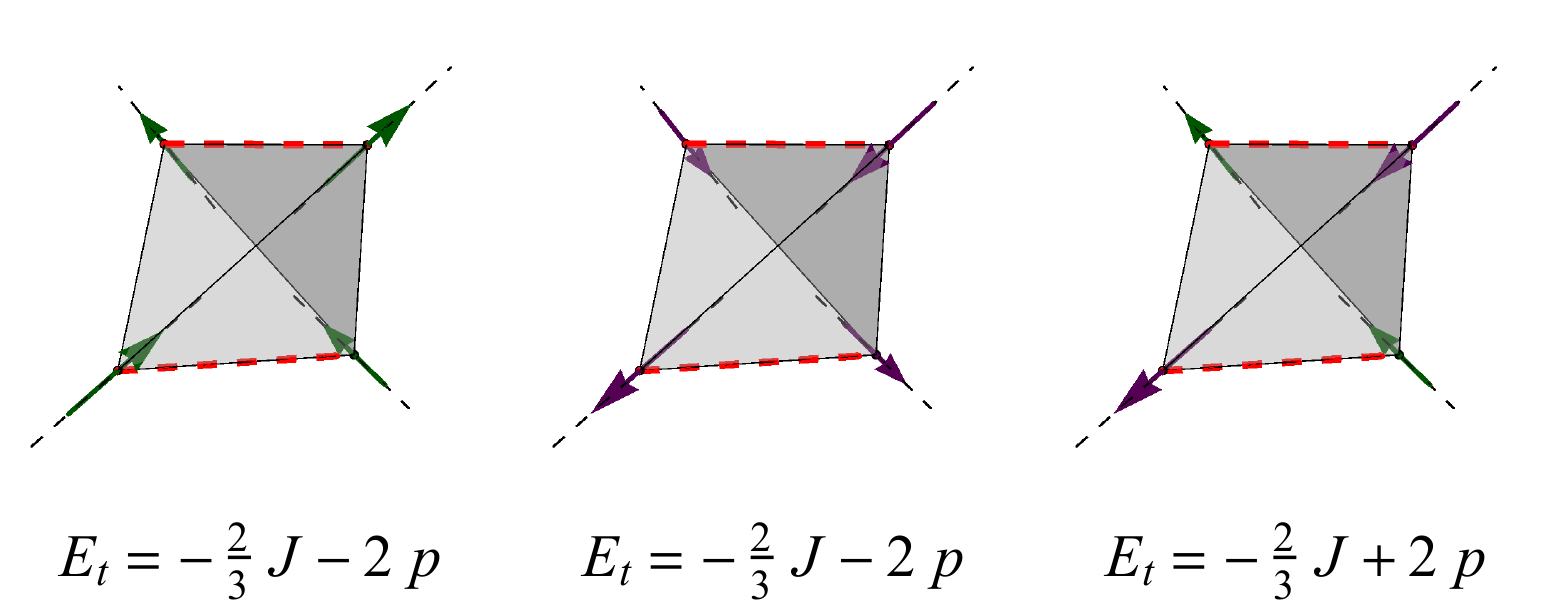}
\caption{Three configurations of a single tetrahedron, and their energy \(E_t\) in the nearest-neighbor Hamiltonian \(\scH\sub{nn}\), \refeq{EqHnn}, with \(J_{ij}\) given in \refsec{SecUniaxialDistortion}. Pairs of spins situated in the same horizontal plane, indicated with dashed (red) lines, have reduced coupling \(J_{ij} = J - 3p\), while others have \(J_{ij} = J\).
All three configurations obey the ice rule, having two spins pointing in and two pointing out. The first two are lowest-energy configurations for a single tetrahedron (since the antiferromagnetically aligned pairs are those with reduced coupling), while the one on the right is one of the remaining four ice-rule configurations whose energy is higher by \(4p\).
Excitations above the ground state are described by strings of spins flipped relative to a fully polarized configuration, and increase the energy by \(4p\) per tetrahedron. When a pair of strings pass through the same tetrahedron, all four spins are inverted and the energy is again minimized; the strings therefore feel an attractive interaction.
\label{FigTetrConfigs}}
\end{figure}

To split the energy of the six ice-rule states on a given tetrahedron requires breaking the cubic symmetry of the pyrochlore lattice. Following \refcite{JaubertPressure}, we consider an explicit uniaxial symmetry breaking, with $J_{ij} = J - 3p$ ($J > 3p > 0$) for pairs of spins whose relative displacement lies in the $(001)$ plane and $J_{ij} = J$ for all others. (Such a perturbation could be effected in experiment by application of uniaxial pressure.\cite{JaubertPressure}) As illustrated in \reffig{FigTetrConfigs}, the result is to favor the two configurations where the total (vector) spin of the tetrahedron is along the $[001]$ axis, whose energy is reduced by $4p$ compared to the other four. In contrast to the case of an applied field,\cite{Jaubert,SpinIce100} an Ising symmetry remains; there are two degenerate lowest-energy states, with all spins on all tetrahedra maximally polarized, consistent with the local easy axes, either along (``up'') or against (``down'') the $[001]$ direction.

We first briefly review the phase structure of the model $\scH\sub{nn}$ with this distortion; readers are referred to \refcites{JaubertPressure,JaubertThesis}\ for more details. For $p \ll T \ll J$, the system remains in the Coulomb phase, while below a critical temperature \(T\sub{c} = 4p(\ln 2)^{-1}\) it becomes a fully polarized ferromagnet. At the transition, the up--down symmetry is broken and the magnetization along the $[001]$ direction, $M_z$, becomes nonzero. While an Ising order parameter can naturally be defined, the transition in the ice-rules limit has quite different properties from the standard Ising universality class. Starting from either of the fully polarized states, the only closed loops are ``strings'' spanning the system in the \([001]\) direction, which cost energy proportional to the (linear) system size \(L\). The transition occurs when the entropy of a single string, also \(\propto L\), outweighs the energy, and so its free energy changes from positive to negative; the string density then increases from zero to nonzero.

As a consequence, the system on the lower-temperature side of the transition is fully polarized, with zero string density, as in the related case of an applied field.\cite{Jaubert,SpinIce100} A crucial distinction in this case is that two strings feel an effective attraction when sharing a tetrahedron (see \reffig{FigTetrConfigs}). At the critical point, this exactly balances the entropy cost of the excluded volume due to the hard-core interactions between strings. In fact, as Jaubert et al.\cite{JaubertPressure}\ have shown, the free energy at $T=T\sub{c}$ as a function of string density is precisely constant (in the thermodynamic limit). Because each string consists of a fixed number of flipped spins relative to the starting configuration, this implies that the free energy is independent of magnetization. As the temperature increases through the transition, the global minimum of $F(M_z)$, which can be interpreted as a Landau function, jumps from $M_z = \pm M\sub{sat}$ to $M_z = 0$. (The resulting discontinuity in the magnetization is illustrated below in \reffig{FigMagnetization0}.) Since all coefficients in the Landau free energy vanish at the transition, this has been referred to as ``infinite-order multicriticality''.\cite{JaubertPressure}

\subsection{Additional interactions}
\label{SecAdditionalInteractions}

Given the magnetization-independent free energy at the transition, it is clear that any perturbation that produces a positive fourth-order coefficient in the Landau function should lead to an intermediate phase with $0 < \lvert M_z\rvert < M\sub{sat}$. While this argument does not provide a prescription for constructing appropriate perturbations, one expects on general grounds that a sufficiently long-ranged four-spin interaction will have this effect. (As will also be demonstrated, a quartic coefficient with opposite sign should lead to a first-order transition.)

\begin{figure}
\putinscaledfigure{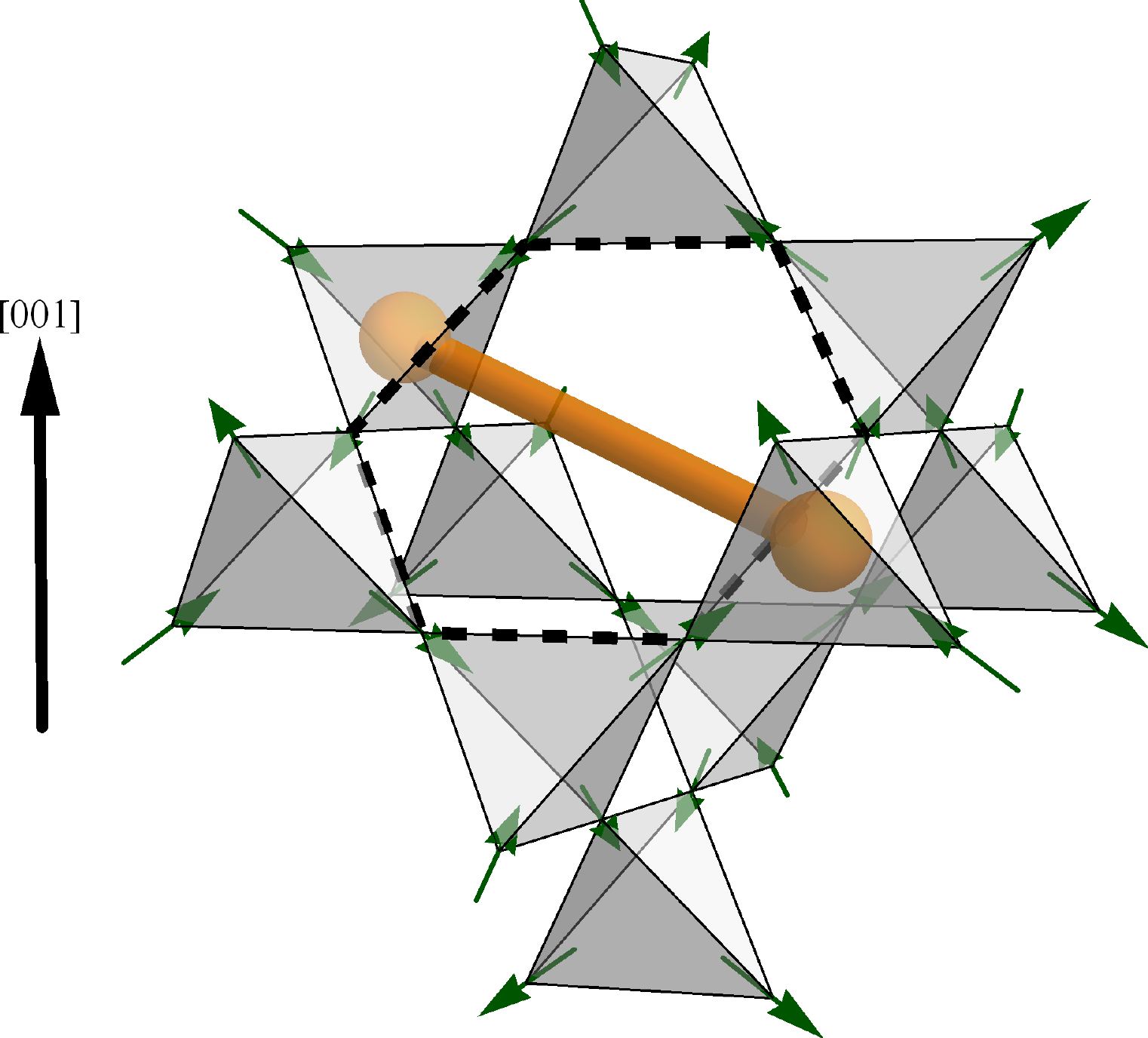}
\caption{Illustration of the four-spin interaction \(\scH\sub{4s}\) added to the model to stabilize the ferromagnetic Coulomb phase. The arrows represent spins on the sites of a pyrochlore lattice, a network of corner-sharing tetrahedra. (This configuration obeys the ice rule, with two spins pointing into and two pointing out of each tetrahedron.)
The additional interaction couples pairs of tetrahedra at opposite sides of hexagons; one such pair and its hexagon are highlighted.
\label{FigPyrochlore}}
\end{figure}

As we detail in the following, MC results in fact demonstrate that it is sufficient to add a four-spin interaction acting between tetrahedra on opposite sides of a hexagonal loop, as illustrated in \reffig{FigPyrochlore}.
The perturbation used throughout this work can be written explicitly as
\beq{EqFourSpin1}
\scH\sub{4s} = V\sub{4} \sum_{\{t t'\}} [\Theta_+(\boldsymbol{S}_t,\boldsymbol{S}_{t'}) + \Theta_-(\boldsymbol{S}_t,\boldsymbol{S}_{t'})]
\punc{,}
\eeq
where
\beq{EqDefineTheta}
\Theta_{\pm}(\Sv,\Sv') = \begin{cases}
1&\text{if \(\Sv=\Sv'=\pm \frac{4}{\sqrt{3}}\hat{\boldsymbol{z}}\)}\\
0&\text{otherwise}
\end{cases}
\eeq
and \(\Sv_t \equiv \sum_{i\in t}\Sv_i\) is the total (vector) spin on tetrahedron $t$. The sum in \refeq{EqFourSpin1} is over pairs of tetrahedra $\{t t'\}$ across a hexagon (see \reffig{FigPyrochlore}), and the summand is one if both tetrahedra have all spins polarized in the same vertical direction, and zero otherwise. (Note that, while this expression apparently involves eight spins, it is equivalent to a four-spin interaction under projection into the ice-rule states. This form of the interaction is partly motivated by the quantum mapping, described in the Appendix.)

Regarding the choice of \(\scH\sub{4s}\), it is not the goal of this work to classify the various types of interactions according to whether they produce a ferromagnetic Coulomb phase, and we are not aware of a general argument that would allow for such a classification.\cite{FootnoteOtherInteractions} (The search for appropriate interactions is in any case better informed by experimental evidence about which interactions occur in particular materials.) Rather, the goal here is to study a particular case where such a phase is known to exist, and elucidate those properties of the phase and its transitions that are expected to be universal, or at least qualitatively generic.

\begin{figure}
\putinscaledfigure{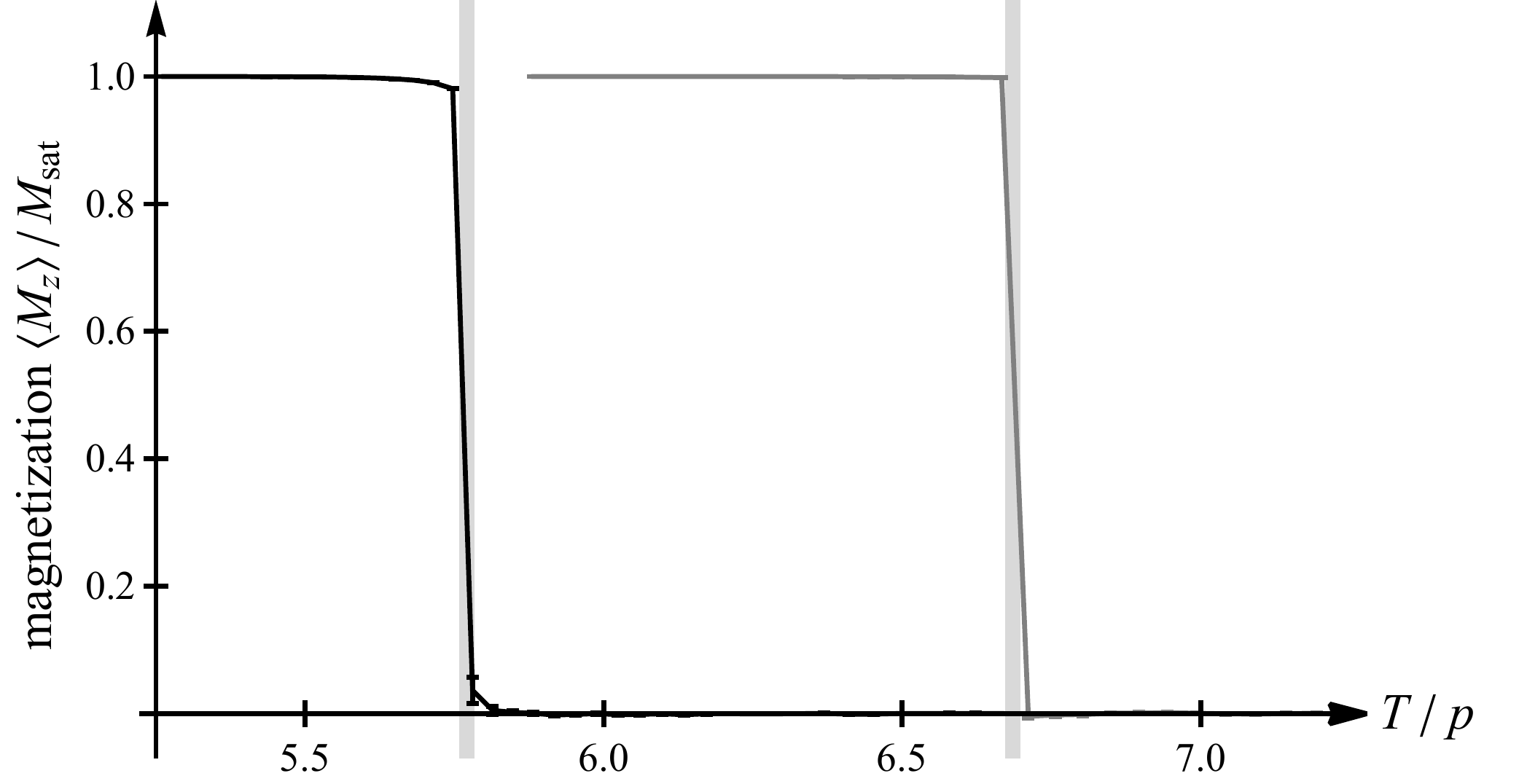}
\caption{Magnetization versus temperature, for fixed $V_4/T = 0$ (left) and $V_4/T = -0.01$ (right), and $L=24$ ($N\sub{s}=16L^3\simeq2\times 10^5$ spins). In both cases, there is a jump from saturation to zero magnetization, at a transition temperature indicated with a vertical line.
For each temperature, the spontaneous magnetization is found by applying a weak field along the $z$ direction and extrapolating to zero field.
\label{FigMagnetization0}}
\end{figure}
\begin{figure}
\putinscaledfigure{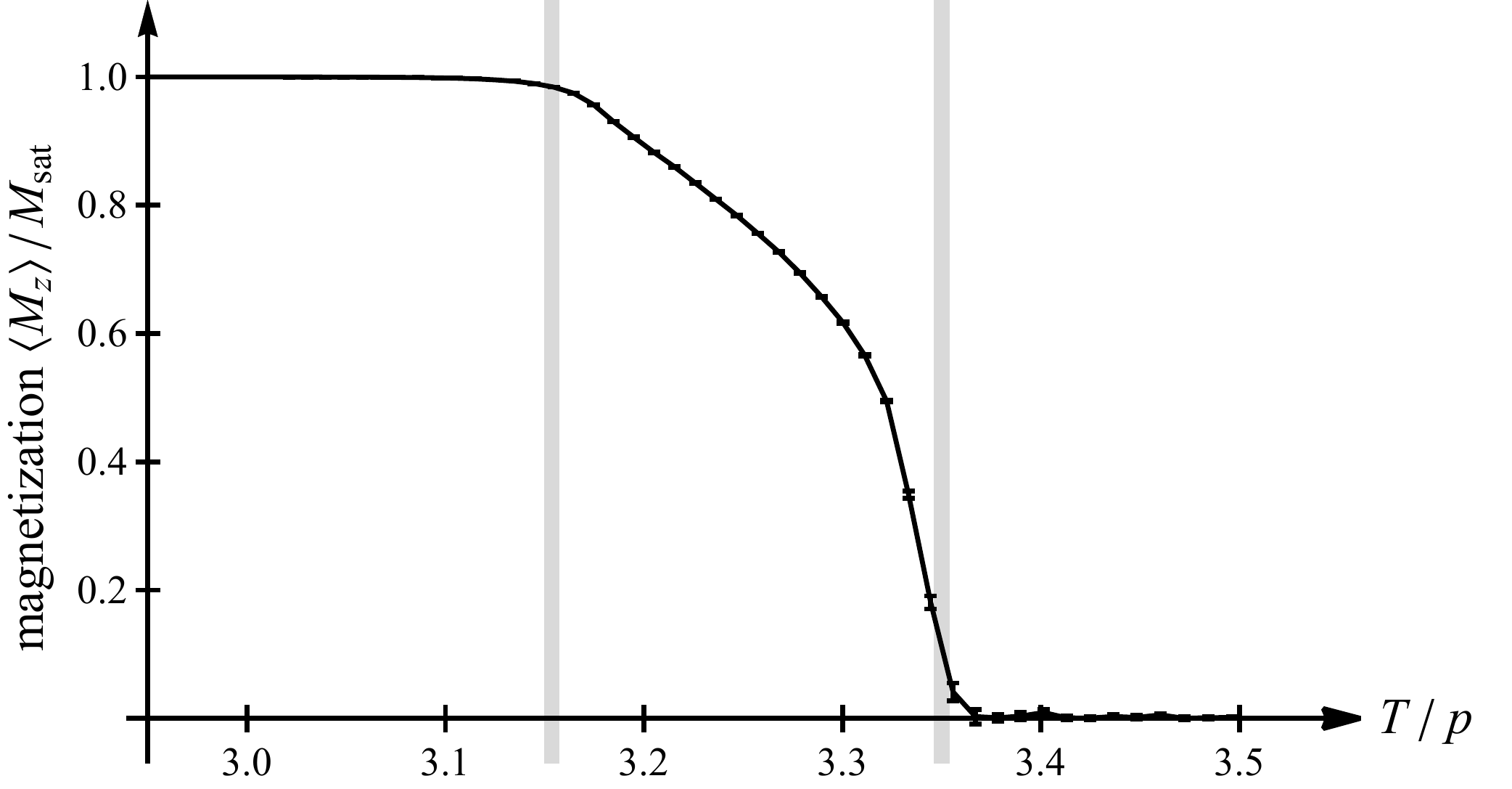}
\caption{Magnetization versus temperature, for fixed $V_4/T = 0.05$ and system size $L=24$.
The vertical lines at $T/p \simeq 3.15$ and $3.35$ indicate positions of phase transitions, determined as described in \refsecand{SecLowerTemperatureTransition}{SecHigherTemperatureTransition} respectively. Below the lower-temperature transition, the magnetization remains at its saturation value, apart from small finite-size corrections, while above the higher-temperature transition, it vanishes. The intermediate phase is a ferromagnet with nonzero and continuously varying magnetization.
\label{FigMagnetization}}
\end{figure}
Plots of the magnetization as a function of temperature, for $V_4$ positive, negative, and zero, are shown in \reffigand{FigMagnetization0}{FigMagnetization}. These results were produced using MC simulations based on a directed-loop algorithm.\cite{Barkema,Sandvik} The lattice consists of $L\times L \times L$ cubic unit cells, each containing $4$ tetrahedra of each orientation, and hence $16$ spins. For $V_4 \le 0$, a step is observed in the magnetization, from essentially fully saturated, with small deviations due to finite-size effects,\cite{JaubertPressure} to zero within error bars. This step is accompanied by a single peak in the specific heat (not shown), whose height grows with system volume, indicating a single first-order transition.

\begin{figure}
\putinscaledfigure{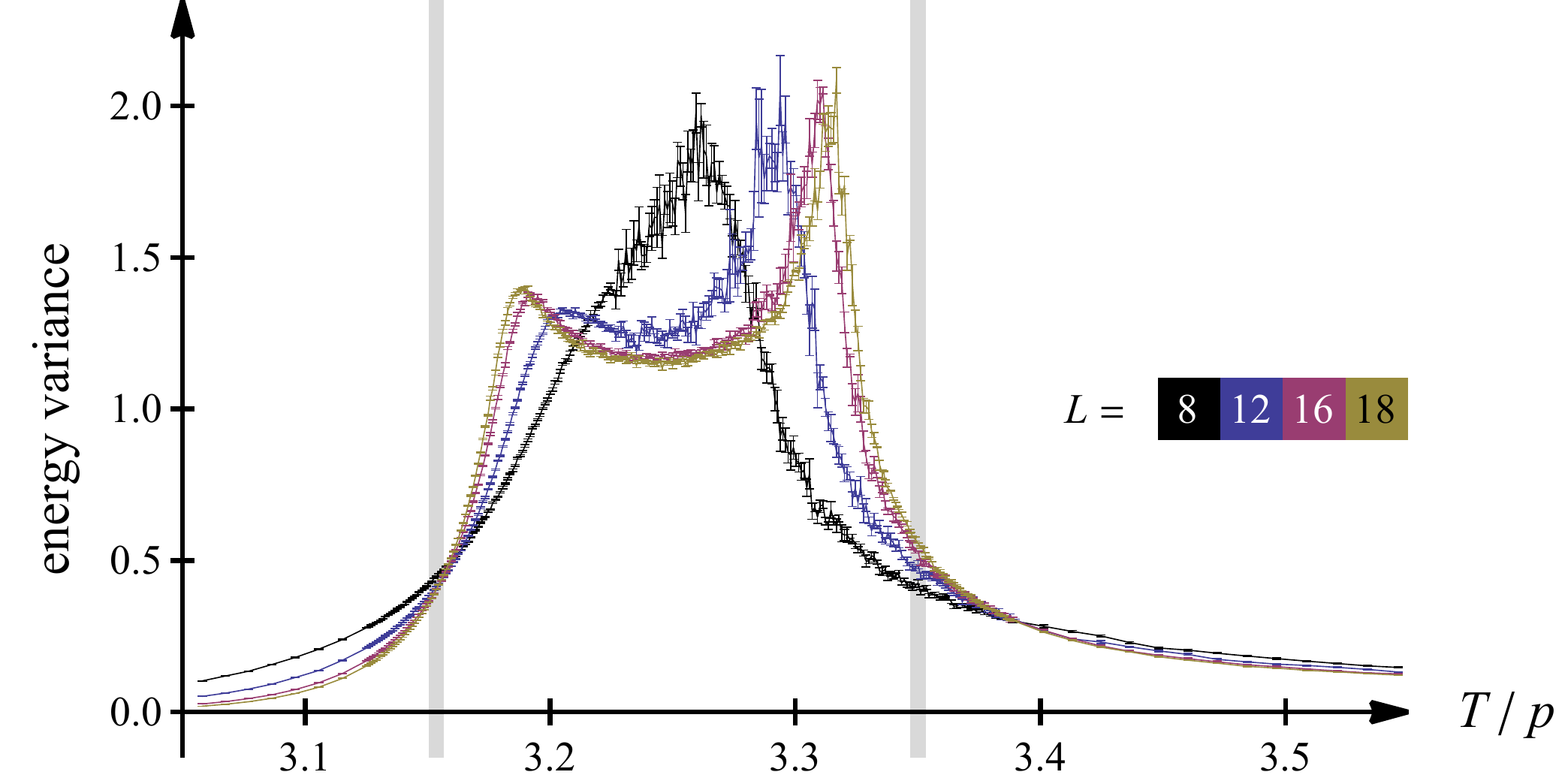}
\caption{Variance of energy, $(\langle E^2 \rangle - \langle E \rangle^2)/N\sub{s}$, versus temperature for fixed $V_4/T = 0.05$ and various system sizes. The vertical lines indicate the positions of phase transitions in the thermodynamic limit (determined by other means). The double-peak structure, with peak heights at most weakly diverging with $L$, is consistent with a pair of continuous transitions.
\label{FigEnergyVariance}}
\end{figure}

By contrast, when $V_4 > 0$ (\reffig{FigMagnetization}), there are clearly three distinct regimes as the temperature $T$ is lowered. The high-temperature phase is paramagnetic, with $\Mv = \zerov$, and is the usual Coulomb phase observed at $T \ll J$ in spin ice. The magnetization first becomes nonzero at $\Tch$ before reaching its saturation value at $\Tcl$. While the system is ferromagnetic for all $T < \Tch$, it is a saturated ferromagnet, with $M_z = \pm M\sub{sat}$, only below $T < \Tcl$. As shown in \reffig{FigEnergyVariance}, the variance of the energy (proportional to the specific heat) in this case displays two peaks, both at most weakly diverging with $L$, consistent with a pair of continuous transitions.

\begin{figure*}
\putinscaledwidefigure{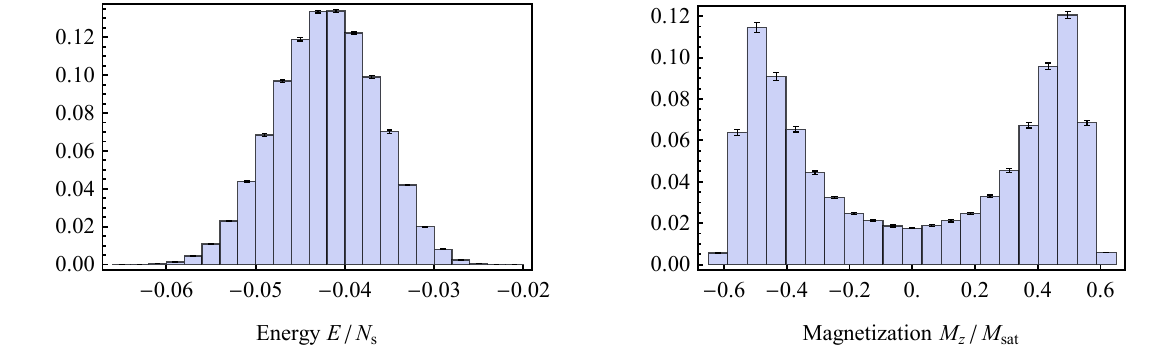}
\vspace{-0.25in}
\caption{Energy and magnetization histograms for $L=16$, $V_4/T = 0.05$, and $T/p = 3.31$ (near the higher-temperature transition). The unimodal energy distribution is indicative of a continuous transition, while the two peaks in the magnetization distribution show that this transition is associated with magnetic ordering and breaking of spin-reversal symmetry.
\label{FigHistogramsPlot}}
\vspace{0.25in}
\end{figure*}

\begin{figure*}
\putinscaledwidefigure{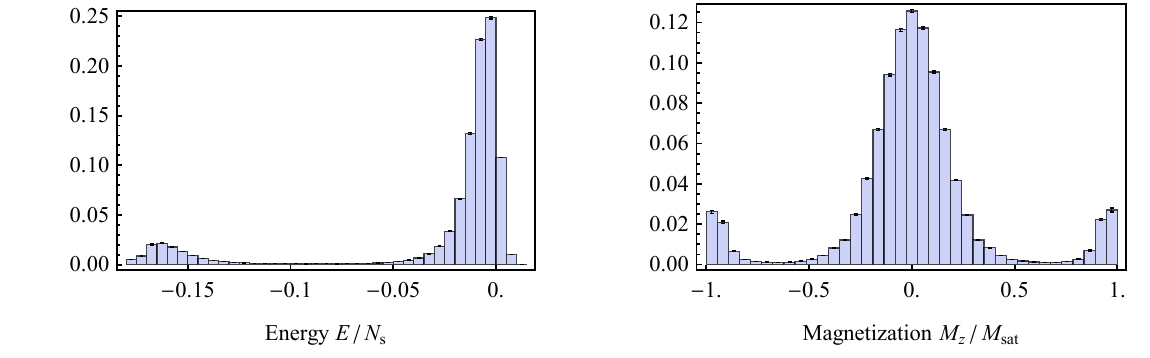}
\vspace{-0.25in}
\caption{Energy and magnetization histograms for $L=8$, $V_4/T = -0.01$, and $T/p = 6.76$. In this case, the transition is strongly first-order, as indicated by the bimodal energy distribution, and occurs directly between the saturated ferromagnet ($M_z = \pm M\sub{sat}$) and the paramagnet ($M_z = 0$). (A small system size is necessary to observe distributions with multiple peaks; for larger $L$ the competing states are metastable.)
\label{FigHistogramsPlot2}}
\vspace{0.25in}
\end{figure*}

\reffig{FigHistogramsPlot} shows histograms of the energy and magnetization for $L=16$, $V_4/T = 0.05$, and $T/p=3.31$, near the higher-temperature peak of the energy variance. The unimodal structure of the energy distribution confirms the continuous nature of the transition, while the two peaks of the magnetization indicate that this is a symmetry-breaking transition into a state with nonzero but unsaturated magnetization.
This should be contrasted with the case of $V_4 = 0$, where the magnetization histogram is flat at the transition.\cite{JaubertPressure} \reffig{FigHistogramsPlot2} shows the case of $V_4 < 0$, where the transition is of first order, with a bimodal structure in the energy and coexisting peaks in the magnetization distribution, at both $M_z = 0$ and $M_z = \pm M\sub{sat}$.

\section{Intermediate phase}
\label{SecIntermediatePhase}

Having established the presence of a pair of phase transitions when $V_4 > 0$, we now turn to the intermediate phase in the temperature range $\Tcl < T < \Tch$. It will be argued that this phase shares the essential spin-liquid features of the Coulomb phase above $\Tch$, but is distinguished by a nonzero spontaneous magnetization.

The presence of a nonzero but unsaturated magnetization in the intermediate phase is evident from \reffigand{FigMagnetization}{FigHistogramsPlot}. Continuously changing magnetization implies that the magnetic susceptibility is nonzero, and hence that there are fluctuations between different topological (magnetization) sectors. This fact alone distinguishes the intermediate phase from the low-temperature saturated ferromagnet, where the flux stiffness vanishes in the thermodynamic limit, and there are no topological-sector fluctuations.\cite{SpinIceReview,Jaubert2013}

Two phenomena that are characteristic of the Coulomb phase are deconfinement and algebraic spin--spin correlations; these are discussed in turn in the following subsections.

\subsection{Monopole distribution function}

A single tetrahedron at which the ice rule is broken (i.e., where the number of spins pointing in and out differs) corresponds to a monopole in the continuous vector field $\Bv(\rv)$. Such defects are rare for $T \ll J$, and, at least as a first approximation, we treat the density of thermally excited monopoles as vanishing.

It is useful to consider, however, the introduction of a single pair of oppositely charged monopoles into an otherwise defect-free background. The effective interaction between the pair, induced by the fluctuations of the surrounding spins, allows one to distinguish spin-liquid phases from others such as the saturated ferromagnet. In the Coulomb phase, the monopoles are subject to an effective Coulomb interaction, with a finite limit for large separation. The saturated ferromagnet is, by contrast, a confining phase, in which the free energy of a pair of monopoles grows without bound as their separation increases.\cite{SpinIceReview,MonopoleScalingPRL,MonopoleScalingPRB}

To determine directly whether monopoles are deconfined, one can define the monopole distribution function $G\sub{m}(\rv_+,\rv_-)$ as the partition function calculated in the presence of a pair of monopoles of opposite charge at $\rv_{\pm}$. (More explicitly, the ensemble is constrained so that all tetrahedra obey the ice rule, apart from those at $\rv_{\pm}$, where three spins point out and one points in, and vice versa.) This function, which is related to the effective interaction between monopoles $U\sub{m}$ by $G\sub{m} = \ee^{-U\sub{m}/T}$, has a nonzero limit for infinite separation $\lvert \rv_+ - \rv_-\rvert$ only when monopoles are deconfined. In a confined phase, it instead decays exponentially to zero.

\begin{figure}
\putinscaledfigure{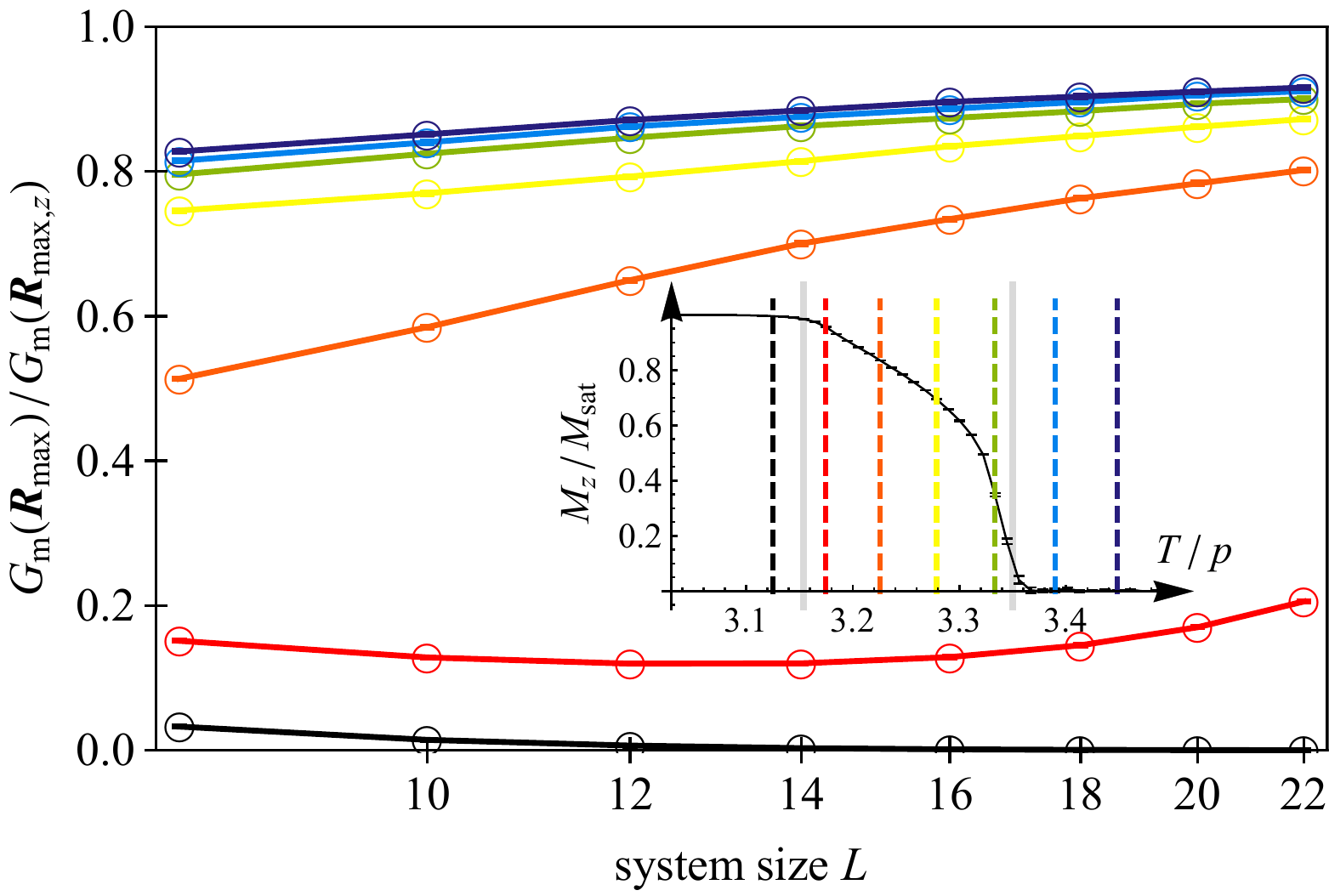}
\caption{Ratio of monopole distribution function $G\sub{m}$ evaluated at the maximum displacement in the lattice, $\Rv\sub{max}$, and at the maximum displacement along the \(z\) direction, $\Rv\sub{max,\(z\)}$. The curves (from bottom to top) have $T/p$ corresponding to the dashed lines (from left to right) in the inset, which reproduces the magnetization  curve of \reffig{FigMagnetization}.
The ratio approaches unity with increasing system size for all temperatures except the lowest, which is within the low-temperature confining phase. No qualitative change is seen across the higher-temperature transition, indicating that the two higher-temperature phases both exhibit deconfinement of monopoles.
Explicitly, the curves have, from bottom to top, $T/p = 3.125$ (black), $3.175$ (red), $3.226$ (orange), $3.279$ (yellow), $3.333$ (green), $3.390$ (light blue), and $3.448$ (dark blue). In all cases, $V_4/T = 0.05$ is fixed.
\label{FigMDF}}
\end{figure}
In a finite system, these asymptotic behaviors are observed only for separations much less than the system size $L$. Finite-size effects can be controlled by fixing the ratio $\lvert \rv_+ - \rv_-\rvert/L$ and observing the scaling with $L$. \reffig{FigMDF} shows the ratio of the monopole distribution function calculated at $\Rv\sub{max}$, the largest displacement possible for $L^3$ cubic unit cells ($L$ even) with periodic boundaries, and at $\Rv\sub{max,\(z\)}$, the maximum separation along the \(z\) direction (\(\lvert\Rv\sub{max}\rvert = \sqrt{3}\lvert\Rv\sub{max,\(z\)}\rvert\)). The ratio approaches unity with increasing system size for all $T > \Tcl$, while it decays to zero below $\Tcl$, indicating confinement. No qualitative difference is seen when crossing the higher-temperature phase boundary at $\Tch$, demonstrating that the intermediate phase, in common with the standard Coulomb phase above $\Tch$, exhibits deconfinement of monopoles.

\begin{figure}
\putinscaledfigure{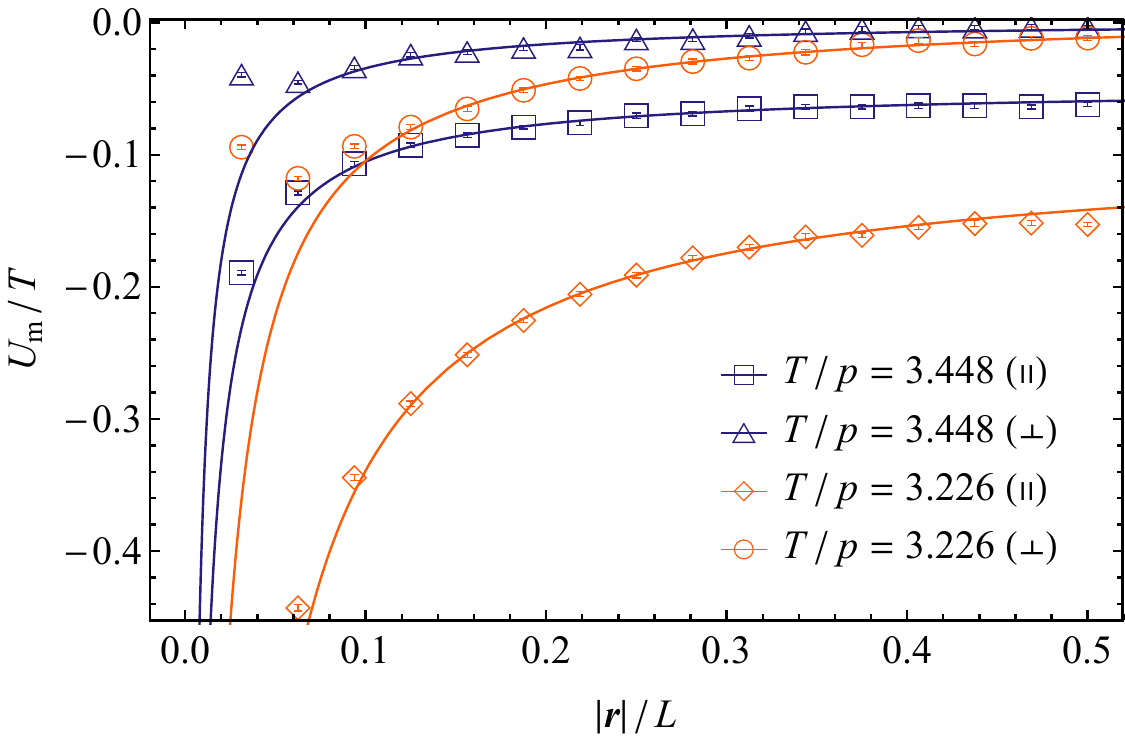}
\caption{Effective (dimensionless) interaction \(U\sub{m}(\rv)/T = -\ln G\sub{m}(\rv)\) between monopoles, for fixed system size \(L = 32\). For each temperature \(T\), the interaction is plotted for separations \(\rv\) parallel (\(\parallel\)) and perpendicular (\(\perp\)) to the axis of the applied pressure, and the zero of interaction is chosen as \(U\sub{m}(\Rv\sub{max}) = 0\). The lines show least-square fits to the Coulomb form \(a - b/\lvert\rv\rvert\) in the region \(0.1 < \lvert\rv\rvert/L < 0.4\), with different parameters \(a\) and \(b\) for each case. There are deviations from the fit at large separation, because of finite-size effects, and at small separation, because of lattice-scale effects and because of the finite range of the additional interactions. (The parameter \(b\) is given by \(b_\parallel = 0.0057\), \(b_\perp = 0.0036\) for \(T/p = 3.448\); and \(b_\parallel = 0.0247\), \(b_\perp = 0.0116\) for \(T/p = 3.226\).)
\label{FigVplot}}
\end{figure}

The form of the effective interaction \(U\sub{m}(\rv) = -T \ln G\sub{m}(\rv)\) is determined by the approach of \(G\sub{m}(\rv)\) to its asymptotic value for large separation \(\rv\). \reffig{FigVplot} shows \(U\sub{m}\) for temperatures within the intermediate phase and above \(\Tch\). In both cases, the interaction is anisotropic, because the spatial symmetry is reduced by the applied pressure (and \(\scH\sub{4s}\)). Up to finite-size effects, the interaction can be fit to the Coulomb form, \(\propto 1/\lvert \rv\rvert\), confirming the identification of the intermediate phase as a Coulomb phase. The effective interaction is stronger parallel to the pressure axis at both temperatures, with larger anisotropy at the lower temperature.

\subsection{Neutron-scattering structure factor}

The most direct experimental signature of the Coulomb phase is the presence of ``pinch points'' in the neutron-scattering structure factor, which reflect the algebraic (dipolar) correlations between the spins.\cite{HenleyReview,Bramwell} These features are clearest in the spin-flip component of polarized neutron-scattering data with incident polarization along $[1\bar{1}0]$.\cite{Fennell} The corresponding structure factor is
\beq{EqDefineSSF}
\scS\sub{SF}(\Qv) = \hat{\eta}_\mu \hat{\eta}_\nu \scS^{\mu\nu}(\Qv)\punc{,}
\eeq
where $\scS^{\mu\nu}(\Qv)$ is the Fourier transform of the two-spin correlation function $\langle S^\mu(\rv) S^\nu(\rv') \rangle$ and
\beq{EqDefineetahv}
\etahv = \frac{\Qv \times \Pv}{\lvert \Qv \rvert \lvert \Pv \rvert}
\eeq
is a unit vector orthogonal to both the wavevector $\Qv$ and the incident neutron polarization $\Pv$.

\begin{figure*}
\putinscaledwidefigure{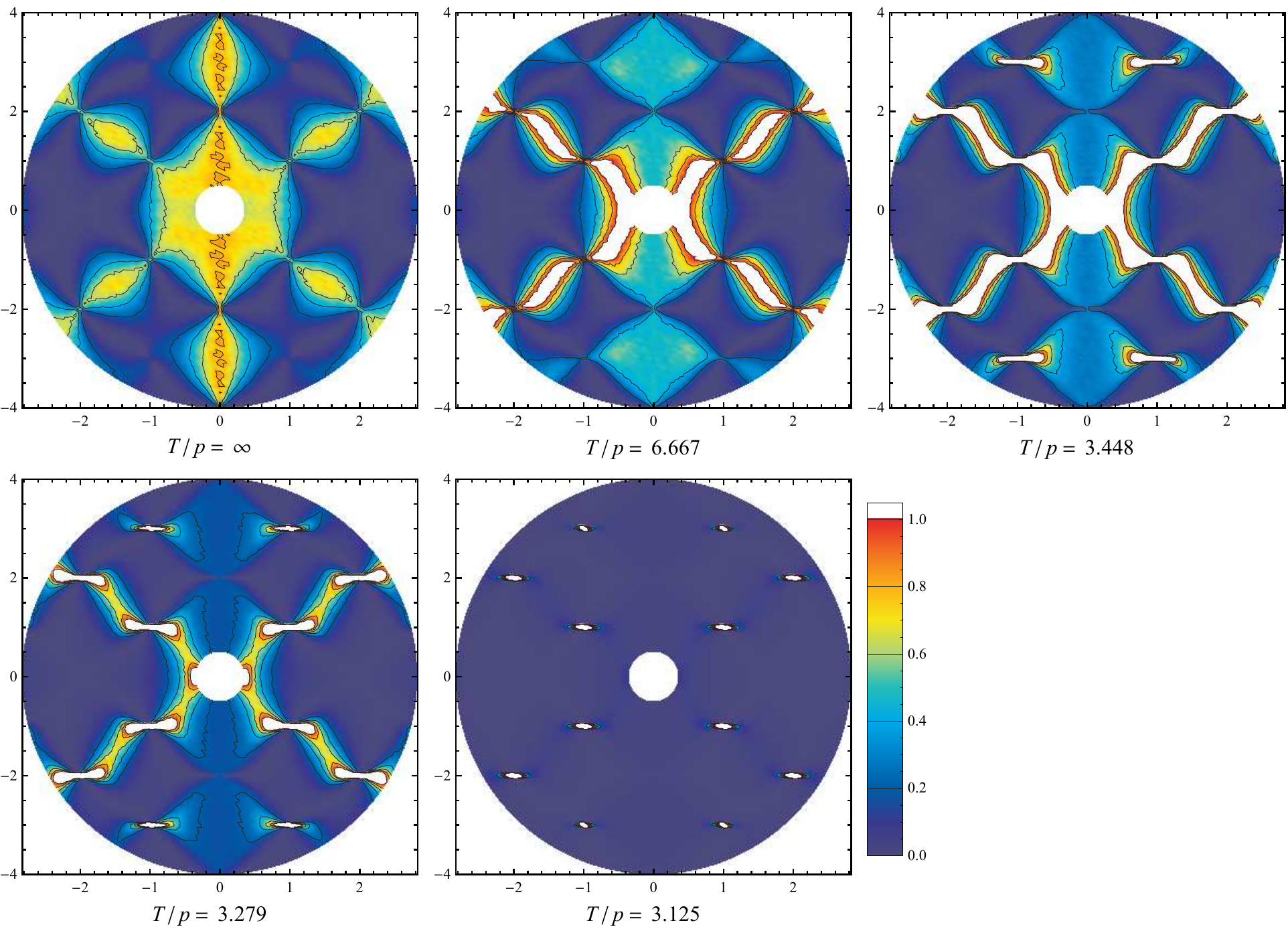}
\caption{Structure factor for (spin-flip) polarized neutron scattering\cite{Fennell} \(\scS\sub{SF}(\Qv)\), defined in \refeq{EqDefineSSF}, for scattering wavevector \(\Qv\) in the \((hh\ell)\) plane and incident polarization \(\Pv\) along \([1\bar{1}0]\).
The first three plots are for temperatures above \(\Tch\), the fourth is in the intermediate phase, \(\Tcl<T<\Tch\), and the last is below \(\Tcl\).
Pinch points, characteristic of the dipolar correlations of the Coulomb phase, are visible at all temperatures but the lowest. In the intermediate phase, there are also Bragg peaks at certain reciprocal lattice vectors, indicating spontaneous magnetization. The system size is $L=16$ and all plots have \(V_4/T = 0.05\). (Wavevectors are measured in units corresponding to the conventional cubic unit cell.)
\label{FigCorrnPlot}}
\end{figure*}

This structure factor is shown in \reffig{FigCorrnPlot}, for \(\Qv\) in the \((hh\ell)\) plane and \(\Pv\) along \([1\bar{1}0]\).
Pinch points are visible for all $T > \Tcl$, with no qualitative change at $\Tch$, showing that the dipolar correlations of the Coulomb phase remain until the lower-temperature transition. In the intermediate phase, they coexist with Bragg peaks at certain reciprocal lattice vectors, indicating ferromagnetic ordering.\cite{FootnoteBragg} The diffuse scattering is completely suppressed in the low-temperature saturated ferromagnet, and only the Bragg peaks remain.

\section{Higher-temperature transition}
\label{SecHigherTemperatureTransition}

The previous sections have established that the phase at $\Tcl < T < \Tch$ is a spin liquid with ferromagnetic order, and that it is connected to the high- and low-temperature phases by continuous transitions. In this section and the following, the critical properties of these two transitions will be addressed in turn, using analytical arguments supported by numerical results.

\subsection{Critical theory}
\label{SecCriticalTheory}

Near the transition, at \(T = \Tch\), between the paramagnetic and ferromagnetic Coulomb phases, the magnetization is far from saturation and so the discrete nature of the spins is presumably not important. Replacing the discrete spins by a continuous vector field\cite{HenleyReview} \(\Bv(\rv)\), the partition function can be expressed as
\beq{EqContinuousAction}
\scZ = \int\DD\Bv \, \delta(\del\cdot\Bv)\, \exp -\!\!\int\!\dd^3\rv\left(\frac{1}{2}\kappa \lvert\Bv\rvert^2 -\frac{1}{2}\alpha B_z^2\right)\punc{.}
\eeq
The coefficient $\kappa$ is the flux stiffness in directions transverse to the applied pressure, while $\alpha > 0$ represents the effect of the uniaxial pressure, enhancing fluctuations along the $z$ direction. (Higher-order terms have been omitted.)

Using a Hubbard--Stratonovich field $\Phi$ to decouple the anisotropy term, this can be replaced by
\begin{multline}
\scZ \propto \int \DD\Phi\int\DD\Bv \, \delta(\del\cdot\Bv)\\\times \exp -\!\!\int\!\dd^3\rv\left(\frac{1}{2\alpha}\Phi^2 + \frac{1}{2}\kappa \lvert\Bv\rvert^2 + \frac{1}{2}\Phi B_z \right)\punc{.}
\end{multline}
The real scalar field $\Phi$ has Ising symmetry and provides an order parameter for the transition, taking a nonzero value in the ferromagnetic phase. Integrating out $\Bv$ induces dipolar interactions for $\Phi$, giving an effective description that is equivalent to that of Ising spins with dipolar couplings. A similar connection between the dipolar correlations in the Coulomb phase and effective dipolar interactions at a critical point has been noted in \refcite{Pickles}.

The 3D Ising transition with dipolar interactions \cite{Larkin,Aharony,Brezin} is at its upper critical dimension, and so shows mean-field critical exponents with logarithmic corrections. In particular, the specific-heat, order-parameter, susceptibility, and correlation-length exponents take the values $\alpha = 0$, $\beta = \frac{1}{2}$, $\gamma = 1$, and $\nu = \frac{1}{2}$, respectively.

It should be noted that scaling remains isotropic at this transition, in the sense that all spatial dimensions scale with the same exponents. For example, the correlation lengths in the directions parallel and perpendicular to the applied pressure diverge with the same exponent $\nu$, though with different (nonuniversal) prefactors. This is in contrast to the anisotropic scaling at the lower-temperature transition (see \refsec{SecLowerTemperatureTransition}).

\subsection{Numerical results}

To determine the critical temperature \(\Tch\) and find values of the exponents, it is convenient to identify a quantity whose scaling dimension vanishes, for which curves with different $L$ coincide at the transition. While the Binder cumulant of the magnetization provides such a quantity for this ordering transition, it is difficult to calculate accurately, as a result of the topological constraints on the magnetization, which suppress fluctuations of the latter.

\begin{figure}
\putinscaledfigure{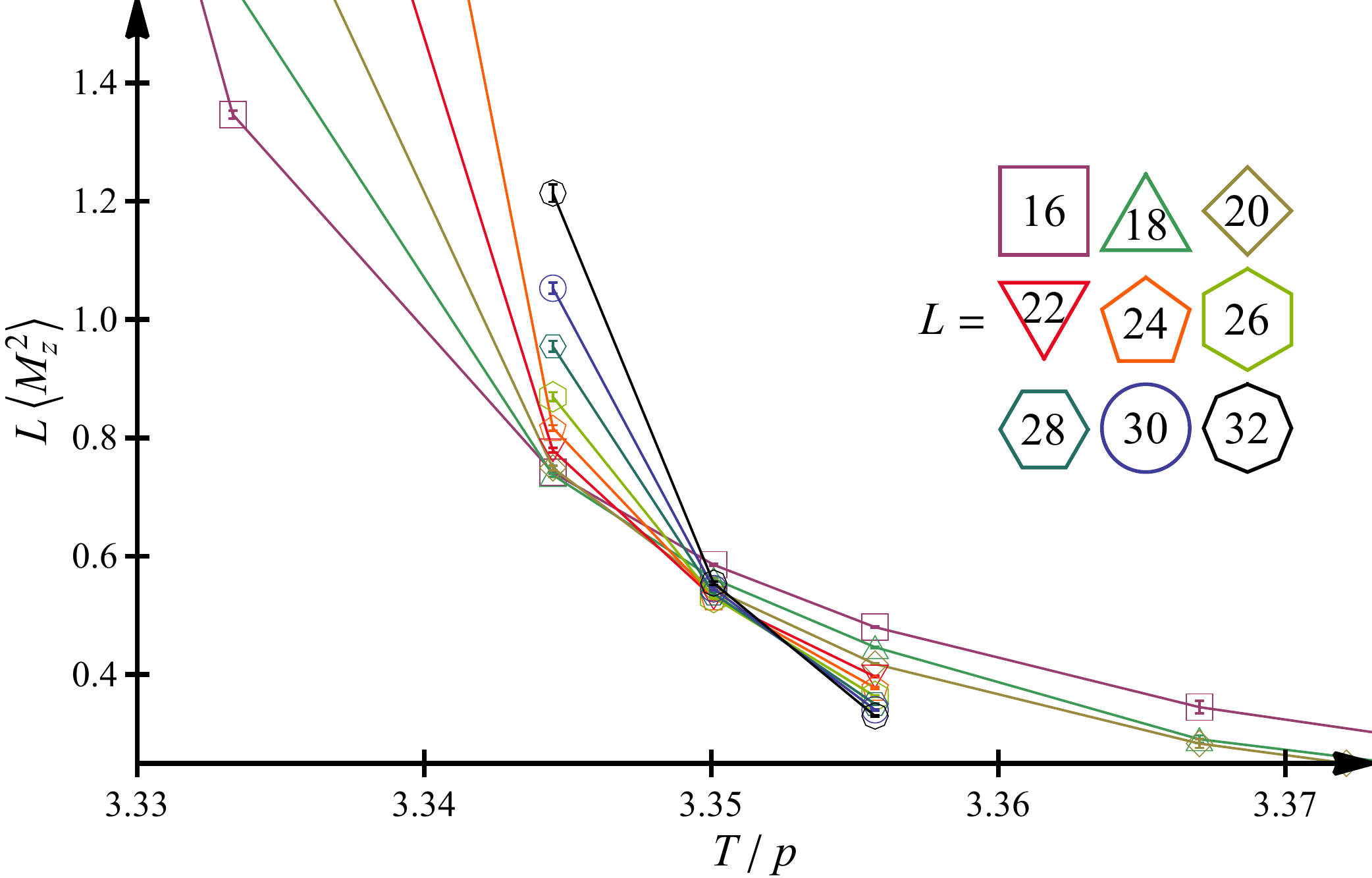}
\caption{Plot of \(L\langle M_z^2\rangle\) versus temperature near the higher-temperature transition, for various system sizes $L$. (In each case $V_4/T = 0.05$ is fixed.) This quantity has vanishing scaling dimension for the mean-field universality class, consistent with the crossing point for large $L$, at $(T/p)\sub{c$>$} = 3.3509(3)$.
\label{FigCrossingPlot}}
\end{figure}

We instead consider the quantity \(L\langle M_z^2\rangle\), which, as a result of the scaling form
\beq{EqScalingMz2}
\langle M_z^2\rangle \approx L^{-d + \gamma/\nu} \Psi\left(L^{1/\nu}\frac{T-\Tch}{\Tch}\right)
\punc{,}
\eeq
where $\Psi$ is a universal function, is expected to have zero scaling dimension for this transition. (This quantity is equal, up to powers of $L$, to the flux stiffness $\Upsilon$, which is not expected to have vanishing scaling dimension at a transition between two spin liquids.) As shown in \reffig{FigCrossingPlot}, \(L\langle M_z^2\rangle\) plotted as a function of $T/p$ indeed has a crossing point for large system sizes. Using the crossings for successive $L$ values, we estimate $(T/p)\sub{c} = 3.3509(3)$ for $V_4/T = 0.05$.

While the observed crossing is consistent with the mean-field exponents, it is also compatible with the Ising universality class, which has\cite{Hasenbusch1999} \(d - \gamma/\nu = 1.0366(8)\). We can go some way to excluding this possibility by calculating the correlation-length exponent $\nu$, which, for the Ising class, takes the value\cite{Hasenbusch1999} $\nu = 0.6298(5)$. \begin{figure}
\putinscaledfigure{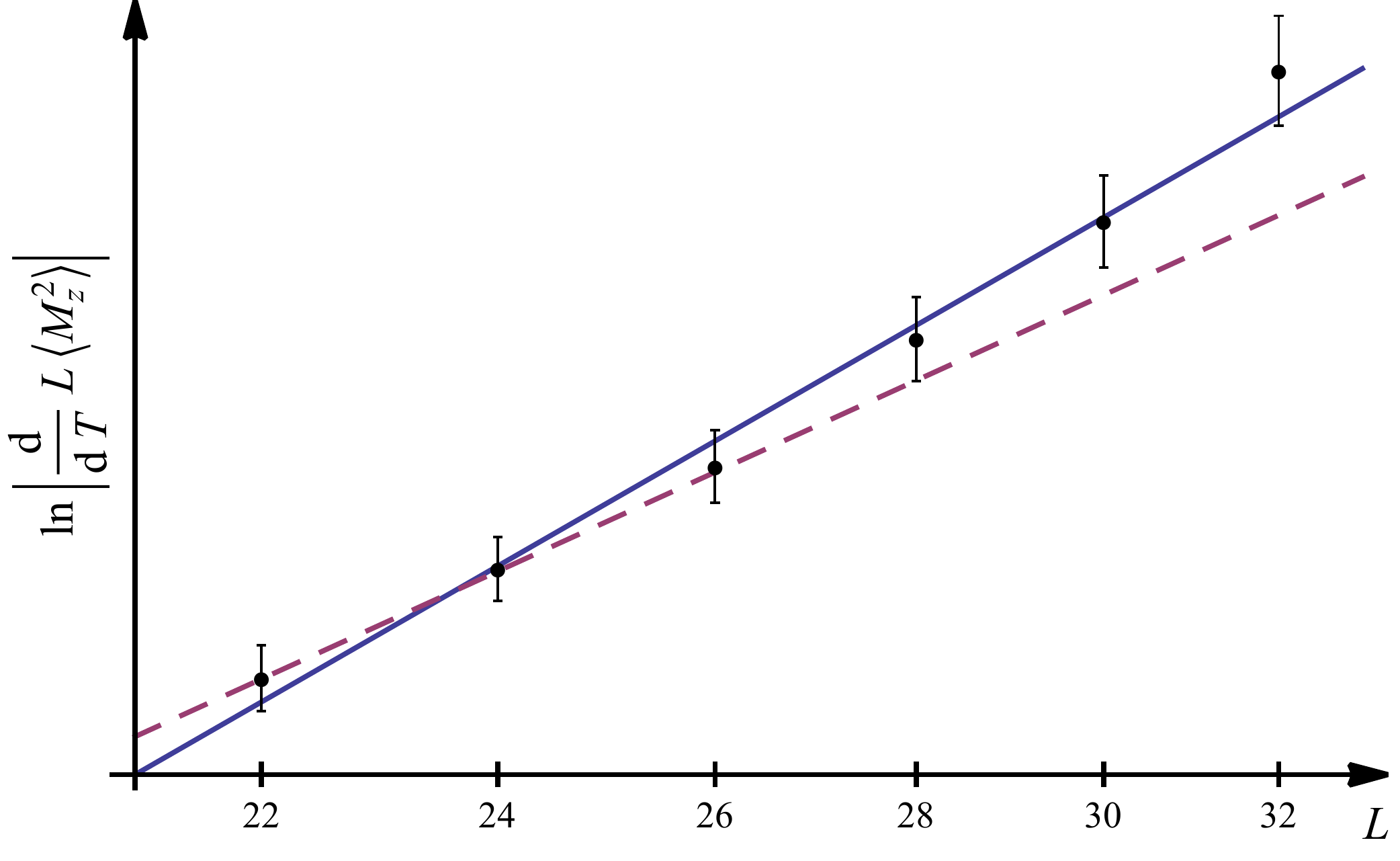}
\caption{Log--log plot of the temperature derivative (in arbitrary units) of \(L\langle M_z^2\rangle\), evaluated at \(T = \Tch\), versus system size \(L\). The slope gives the reciprocal of the correlation-length exponent \(\nu\). The (blue) solid line has \(\nu = \frac{1}{2}\), as expected for the mean-field universality class, while the (purple) dashed line has the Ising value \(\nu = 0.63\).
\label{FigSlopePlot}}
\end{figure}
\reffig{FigSlopePlot} shows the temperature derivative of \(L\langle M_z^2\rangle\) evaluated at \(T = \Tch\), which is expected to scale as
\beq{EqScalingDerivative}
\left.\frac{\dd}{\dd T} L\langle M_z^2\rangle \right\rvert_{T = \Tch} \sim L^{-d+\gamma/\nu+1+1/\nu} = L^{1/\nu}\punc{.}
\eeq
While not conclusive, the results are consistent with Ising-like criticality for smaller system sizes, crossing over to the true mean-field universality class for \(L > 25\).

For the available system sizes, there is no evidence of the expected logarithmic corrections to scaling. We do not consider this to be strong evidence for their absence, however, since much larger systems are often required to observe logarithmic corrections.\cite{Cardy}

\section{Lower-temperature transition}
\label{SecLowerTemperatureTransition}

The lower-temperature transition, at \(T = \Tcl\), separates the ferromagnetic Coulomb phase from a conventional ferromagnet. Because the magnetization is nonzero on both sides, the spin-inversion symmetry of the Hamiltonian is immaterial, and the transition is in the same universality class as the saturation transition in an applied field.\cite{Jaubert,SpinIce100} This is a Kasteleyn transition, which exhibits anisotropic scaling in the directions parallel and perpendicular to the magnetization, with relative scaling exponent $\scz = 2$. The transition is consequently at its upper critical dimension, and so shows mean-field exponents with logarithmic corrections.\cite{Jaubert,MonopoleScalingPRB}

\begin{figure}
\putinscaledfigure{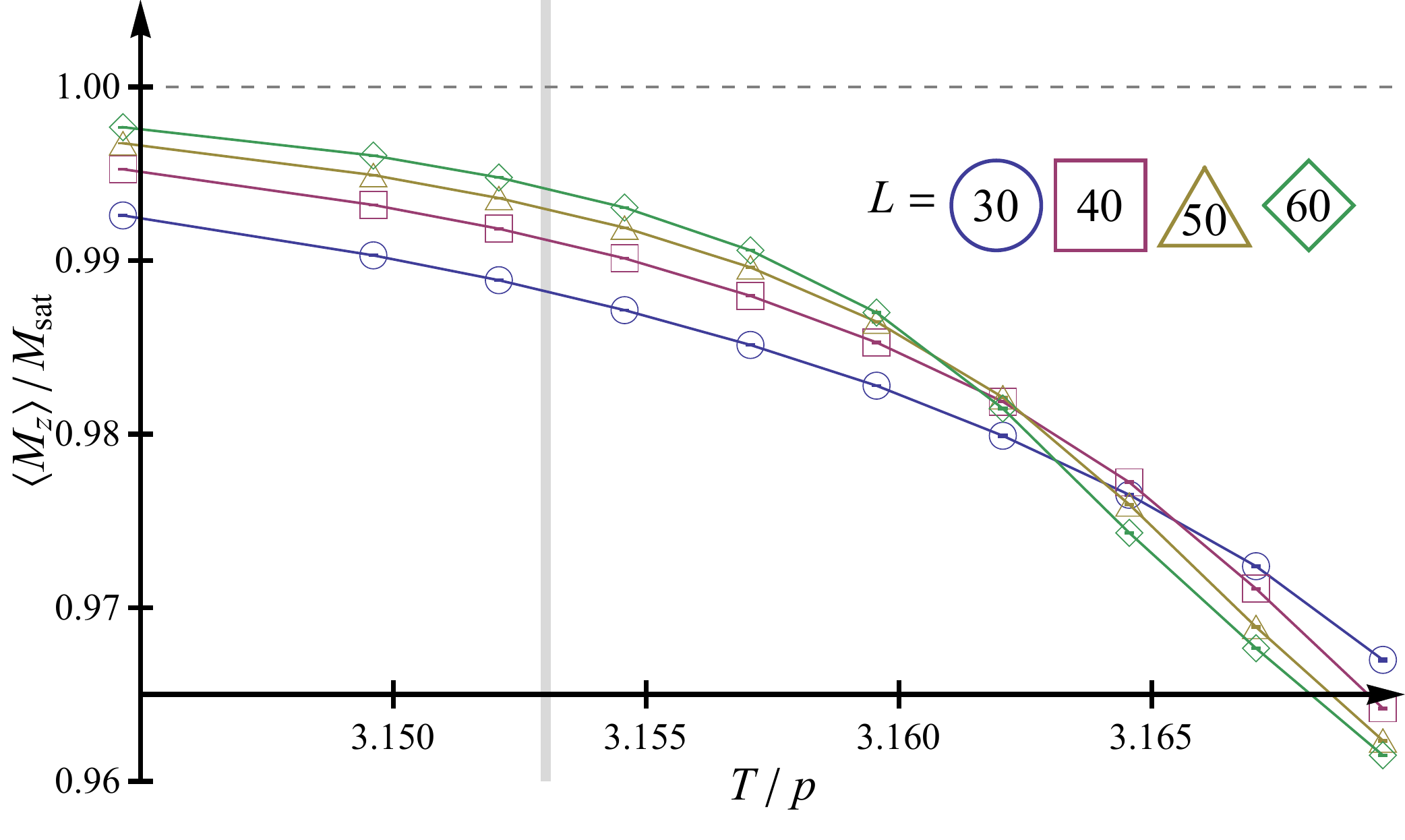}
\caption{Magnetization near the lower-temperature transition at \(\Tcl\), indicated with a vertical line, for large system sizes. As $L$ increases, the magnetization approaches saturation below the transition, and a kink develops at \(\Tcl\) (compare also \reffig{FigMagnetization}, for \(L=24\)). In this case, the MC simulation is run starting from a state with saturated magnetization, \(\langle M_z\rangle = M\sub{sat}\); the low temperatures and large system sizes ensure that ergodicity is broken and the order parameter takes a nonzero value.
\label{FigMagnetizationLarge}}
\end{figure}

The Kasteleyn transition has the distinguishing characteristic that the magnetization is saturated in the low-temperature phase (in the thermodynamic limit), and decreases continuously but nonanalytically across the transition.\cite{Kasteleyn,Jaubert} The magnetization is plotted in \reffig{FigMagnetizationLarge} for large systems near the lower-temperature transition, showing the development of a kink as the system size grows and indicating that the departure from saturation magnetization for \(T < \Tcl\) is a finite-size effect.

\begin{figure}
\putinscaledfigure{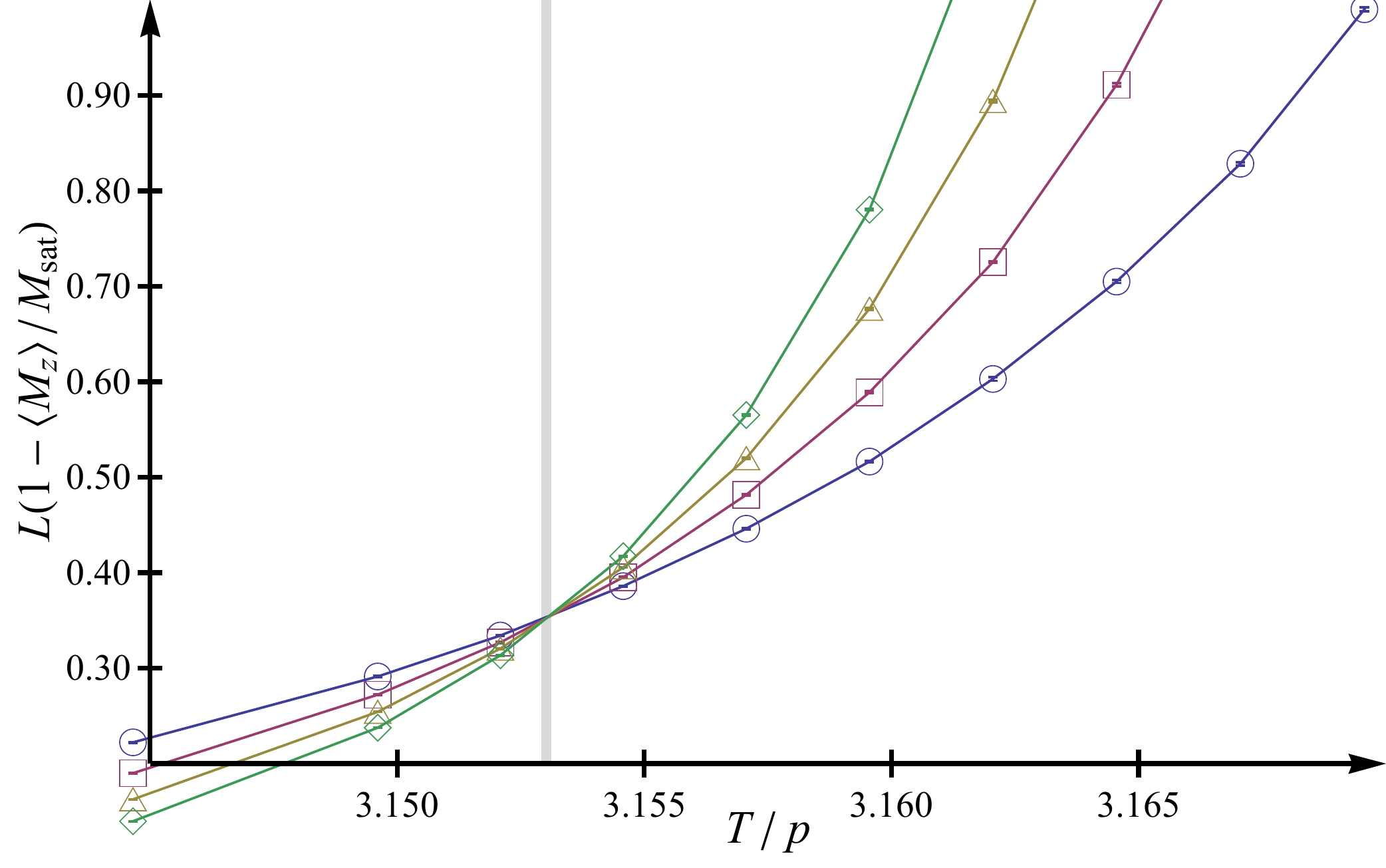}
\caption{Order parameter for the lower-temperature transition, \(1-\langle M_z\rangle/M\sub{sat}\), multiplied by \(L\) and plotted versus temperature for various \(L\) (using the same symbols as \reffig{FigMagnetizationLarge}). This quantity has vanishing scaling dimension at the Kasteleyn transition; a crossing is observed at \((T/p)\sub{c$<$} = 3.15302(9)\).
\label{FigCrossingLarge}}
\end{figure}

Although there is no symmetry breaking at the transition, the quantity \(1-\langle M_z\rangle/M\sub{sat}\) can be identified as an order parameter, taking a nonzero value only on the high-temperature side. The critical theory for the transition\cite{Jaubert,SpinIce100} can be written using a \(\mathrm{U}(1)\)-symmetric complex field \(\psi\), in terms of which the order parameter is given by \(1-\langle M_z\rangle /M\sub{sat} \sim \psi^*\psi\). It follows that the scaling dimension of \(L(1-\langle M_z\rangle/M\sub{sat})\) vanishes,\cite{MonopoleScalingPRB} and so a crossing point is expected when this quantity is plotted for different \(L\). This crossing is shown in \reffig{FigCrossingLarge}, enabling \((T/p)\sub{c$<$} = 3.15302(9)\) to be found and providing confirmation of the Kasteleyn universality class.

Previous MC simulations of spin ice in an applied field\cite{Jaubert,MonopoleScalingPRB} have shown that logarithmic corrections are visible for \(L \gtrsim 100\). In the present case, the further-neighbor interactions in the model make the MC simulations more computationally demanding, and only mean-field behavior is observed at accessible system sizes.

The Kasteleyn transition occurs when the magnetization first deviates from saturation, which occurs through the appearance of strings of spins flipped relative to the fully polarized state.\cite{Jaubert} One can therefore determine the exact transition temperature by considering the free energy of a single string, and finding the point where this becomes negative.\cite{Jaubert,JaubertPressure} When \(V_4 = 0\), such a string contributes free energy of \(\Delta f = 4p - T \ln 2\) per unit length, where the second term reflects the entropy associated with the possible paths. Including the four-spin interaction \(V_4\) modifies this to
\beq{EqStringFreeEnergy}
\Delta f = 4p - T \ln \left(\ee^{12V_4/T} + \ee^{11V_4/T}\right)\punc{,}
\eeq
because the two possible routes for the string (following a \(\langle 011 \rangle\) chain or otherwise) have different energies. The Kasteleyn transition occurs when \(\Delta f = 0\), giving \((T/p)\sub{c$<$} = 3.15343\) for \(V_4/T = 0.05\). The discrepancy with the numerical results, which is small in absolute terms but several times the estimated statistical error, may result from logarithmic corrections to the scaling form for \(\langle M_z\rangle\).

Finally, it should be noted that, while both the higher- and lower-temperature transitions are at their upper critical dimensions, and hence have rational exponents, they otherwise have quite different properties. The Kasteleyn transition has anistropic scaling in directions parallel and perpendicular to the magnetization, while at the higher-temperature transition the system scales isotropically. A second distinction is that the magnetization is the critical field for the higher-temperature transition, while it is related to a bilinear of the critical field \(\psi\) for the Kasteleyn transition.

\section{Discussion}
\label{SecDiscussion}

This work has studied a nearest-neighbor model of spin ice with uniaxial distortion, which has a ferromagnetic phase at low temperature. Analytical arguments were used to show that an additional four-spin interaction can lead to an intermediate phase with coexisting ferromagnetic order and spin-liquid characteristics; the presence of this ferromagnetic Coulomb phase (FCP) has been established using Monte Carlo simulations.

Many features of the FCP are expected to occur more generally in spin liquids, both classical and quantum, with magnetic order. A clear experimental signature of an ordered Coulomb phase is coexistence of Bragg peaks, indicating magnetic order, with pinch points (see \reffig{FigCorrnPlot}). On the theoretical side, a defining characteristic of spin-liquid phases is fractionalized excitations, such as magnetic monopoles in spin ice and spinons in quantum antiferromagnets; these remain deconfined across the transition into an ordered spin liquid (see \reffig{FigMDF}). Finally, such phase transitions have conventional order parameters, but their critical properties are modified by coupling to the soft modes of the spin liquid (see \refsec{SecHigherTemperatureTransition}).

The analysis here, including the Monte Carlo data, has treated the ice rule (see \refsec{SecModel}) as a strict constraint on configurations of the model. With a nonzero but small density of monopoles (i.e., defects in this constraint), as in the spin-ice compounds at low temperature, one expects most of the results to apply essentially unchanged: While the lower-temperature transition is immediately replaced by a crossover, this remains sharp for small monopole density.\cite{Jaubert} The FCP is no longer qualitatively distinct from a conventional ferromagnet, but there can be a clear regime where the system is effectively described by a classical spin liquid with a small density of monopoles.\cite{MonopoleScalingPRB} The higher-temperature transition remains, but is strictly in the Ising universality class at any nonzero monopole density; as in the case of the cubic dimer model,\cite{Sreejith} however, the critical behavior is strongly affected by the presence of the unconventional critical point at zero monopole density.

\acknowledgments

I am grateful to John Chalker, Ludovic Jaubert, and Michael Levin for helpful discussions. The Monte Carlo simulations used resources provided by the University of Nottingham High-Performance Computing Service.

\appendix*

\section{Quantum mapping}
\label{SecQuantumMapping}

In this Appendix, we consider a model of quantum bosons in two spatial dimensions (2D), which shows closely analogous behavior to the model of spin ice discussed in the body of the paper. In fact, using the general mapping between classical statistical mechanics in 3D and quantum mechanics in 2D, which has previously been applied to phase transitions from CSL phases,\cite{SpinIce100,CubicDimers} one expects the universal features of the phases and transitions to be equivalent in these two models.

The nearest-neighbor model for spin ice, \(\scH\sub{nn}\), can be mapped to a system of hard-core lattice bosons, with spin-reversal symmetry replaced by particle--hole symmetry. The Coulomb phase of the spin model is equivalent to a superfluid,\cite{Jaubert,SpinIce100} while the saturated ferromagnet with \(M_z = \pm M\sub{sat}\) is equivalent to the vacuum and fully-occupied states of the bosonic model, which spontaneously break particle--hole symmetry. The strings of flipped spins that proliferate at the transition (see \refsec{SecUniaxialDistortion}) map to boson world lines (trajectories in space and imaginary time).

An equivalent bosonic model to the Hamiltonian \(\scH\sub{nn}\), displaying infinite-order multicriticality at the transition between these two phases, is given by\cite{RokhsarKotliar}
\beq{EqHamQuantum0}
\scH_0 = -t \sum_{\langle i j \rangle} (b_i^\dagger b_j + b_j^\dagger b_i) - V \sum_{\langle i j \rangle} \left[(n_i - \tfrac{1}{2})(n_j - \tfrac{1}{2}) - \tfrac{1}{4}\right]\punc{,}
\eeq
where \(b_i = \lvert 0 \rangle_i \langle 1 \rvert_i\) and \(n_i = b_i^\dagger b_i = \lvert 1 \rangle_i \langle 1 \rvert_i\) are annihilation and number operators for hard-core bosons. The first term represents tunneling \(t\) between neighboring sites \(\langle ij \rangle\), while the second is an attractive interaction of strength \(V > 0\) between nearest-neighbor bosons, written in a manifestly particle--hole-symmetric form.

Because \(\scH_0\) conserves particle number, the Hilbert space can be divided into sectors of fixed density \(\rho = \langle n_i \rangle\); let \(E\sub{gs}(\rho)\) be the ground-state energy in each. For \(t > V\), the overall ground state occurs in the sector with \(\rho = \frac{1}{2}\), and the system is a particle--hole-symmetric superfluid. For \(t < V\), \(E\sub{gs}\) is instead minimized by either the vacuum, \(\rho = 0\), or the fully-occupied state, \(\rho = 1\). At \(t = V\), the model has a Rokhsar--Kivelson (RK) point,\cite{RokhsarKivelson} at which \(\scH_0\) can be written as a projector,
\beq{EqProjector}
\scH_0 = \frac{1}{2}t\sum_{\langle ij \rangle}(\lvert 1 \rangle_i \lvert 0 \rangle_j - \lvert 0 \rangle_i \lvert 1 \rangle_j)(\langle 1 \rvert_i \langle 0 \rvert_j - \langle 0 \rvert_i \langle 1 \rvert_j)\punc{,}
\eeq
and the ground state in each sector, an equal-amplitude superposition of all configurations,\cite{RokhsarKotliar} has \(E\sub{gs}(\rho) = 0\).
\begin{figure}
\putinscaledfigure{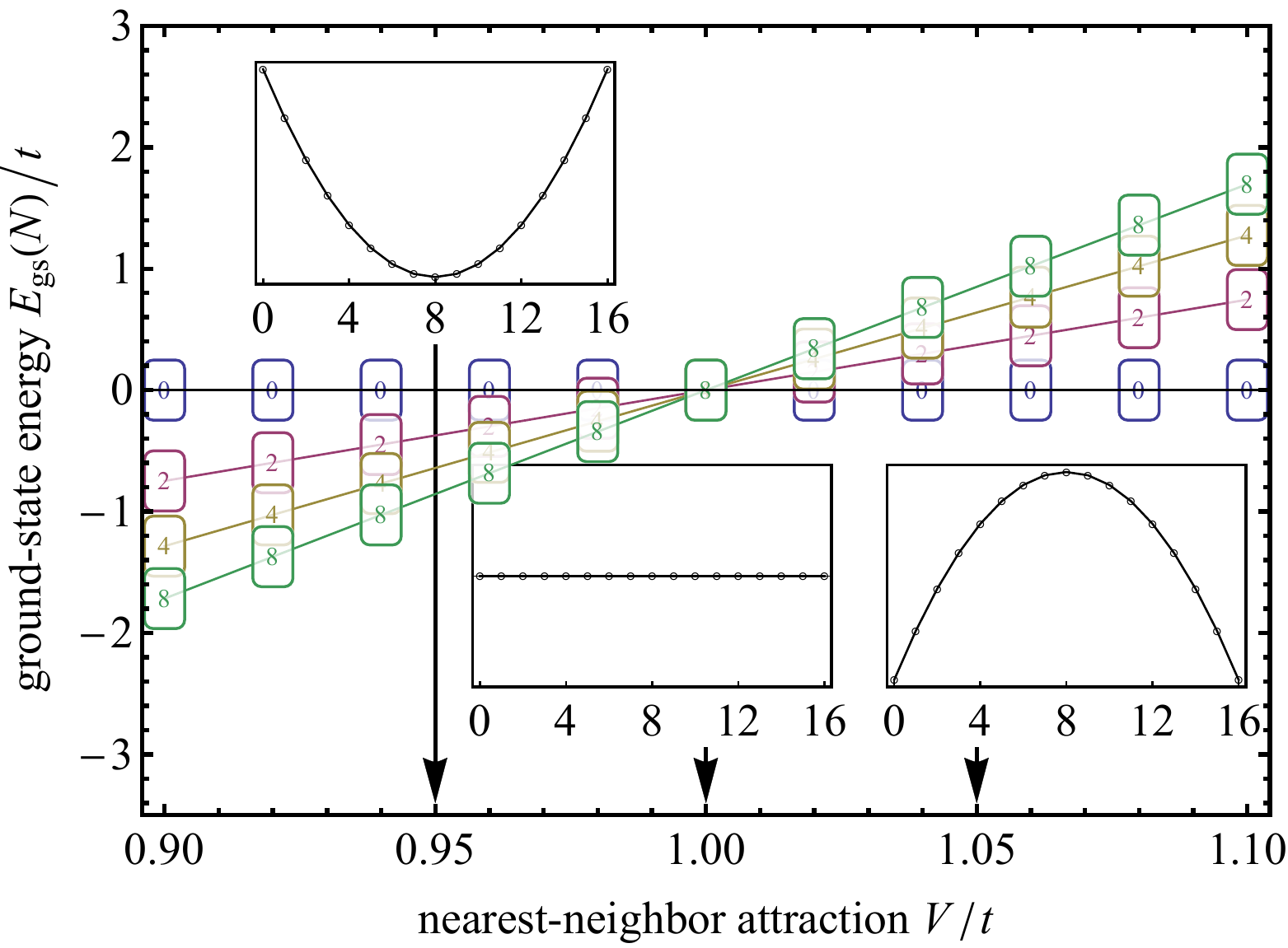}
\caption{Ground-state energy \(E\sub{gs}\), in the Hilbert-space sector with \(N\) particles, for hard-core bosons with particle--hole symmetry, \refeq{EqHamQuantum0}. Results were obtained using exact diagonalization on a \(4\times 4\) square lattice with periodic boundary conditions. The main figure shows \(E\sub{gs}(N)\), for the labeled values of \(N\), versus nearest-neighbor attraction \(V\), both in units of the tunneling strength \(t\). The model has an RK point at $V = t$, at which \(E\sub{gs}\) is independent of \(N\). For \(V<t\) the ground state of the system occurs for half filling, \(N = 8\). For \(V > t\) there are two generate ground states, with boson density zero and one (\(N=0\) and \(N=16\) respectively), which spontaneously break particle--hole symmetry.
The insets show \(E\sub{gs}\) versus \(N\) for fixed values of \(V/t\), indicated by the arrows.
\label{FigEgsPlot0}}
\end{figure}
As illustrated in \reffig{FigEgsPlot0}, which shows results of exact diagonalization (ED) on a small system, this leads to a transition with identical properties to the ordering transition of spin ice under uniaxial pressure, with \(F(M_z)\) replaced by \(E\sub{gs}(\rho)\). (The exact equivalence is established by noting that the transfer matrix for the classical problem can be written as a projector at the transition.\cite{JaubertPressure})

Following similar logic to \refsec{SecAdditionalInteractions}, one expects that quartic interactions between bosons should open up an intermediate phase with density changing continuously between \(\rho = \frac{1}{2}\) and \(\rho = 0\) or \(1\).
\begin{figure}
\putinscaledfigure{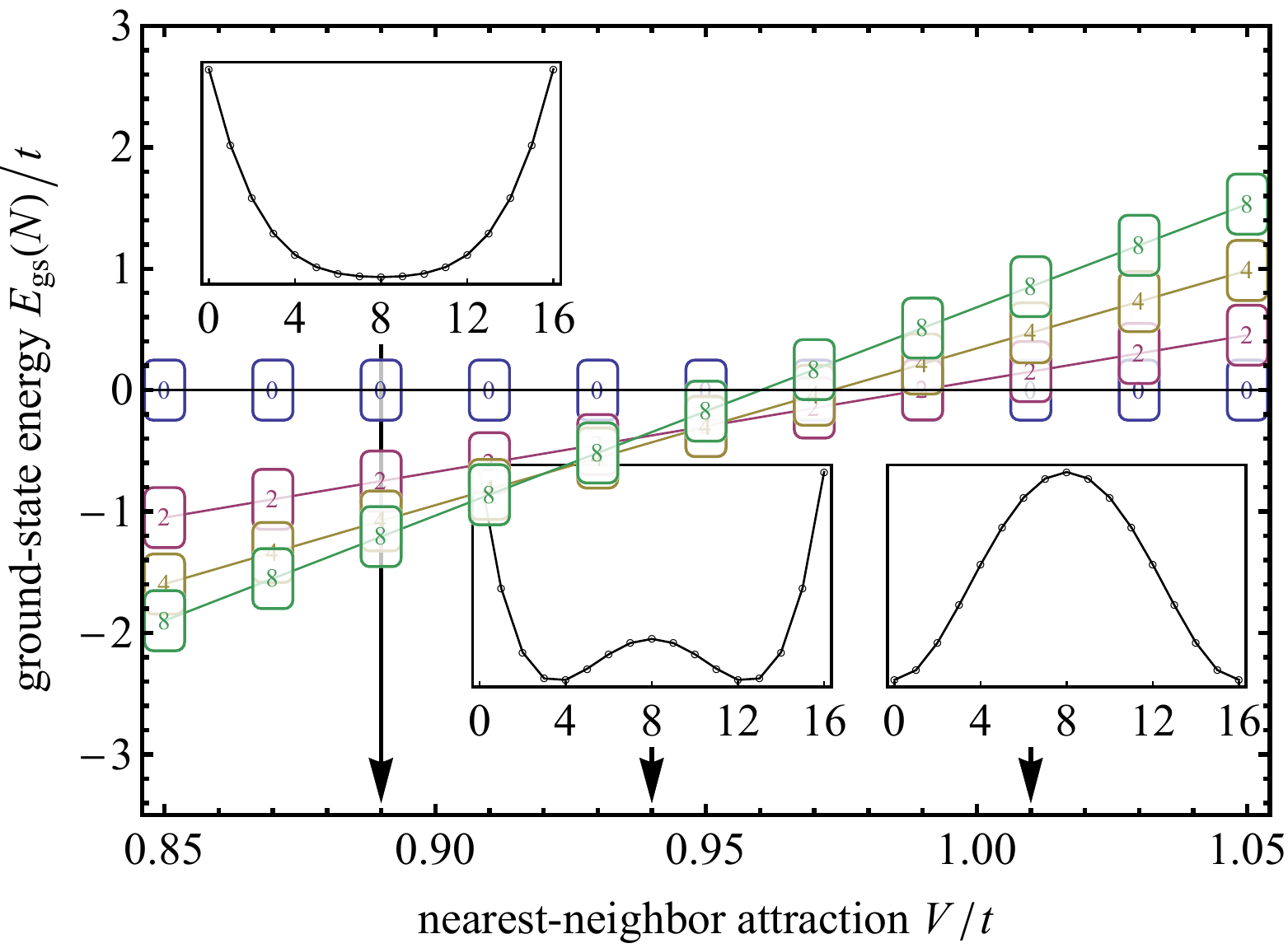}
\caption{Ground-state energy \(E\sub{gs}(N)\) versus nearest-neighbor attraction \(V\), as in \reffig{FigEgsPlot0}, with an additional four-body repulsion \(V_4 = 0.3 t\), \refeq{EqFourBody}. As in the case with \(V_4 = 0\), the ground state is at half filling for \(V \ll t\) and has density either zero or one for \(V \gg t\). In between, there is a regime where the minimum of \(E\sub{gs}\) crosses over between the extremes, stepping through each intermediate density in turn. In the thermodynamic limit, this is expected to become a phase with continuously varying density, separated from small- and large-\(V/t\) phases by continuous quantum phase transitions.
\label{FigEgsPlot}}
\end{figure}
The precise form required is again unclear from these general arguments, but ED results, shown in \reffig{FigEgsPlot}, indicate that it suffices to add a (particle--hole-symmetrized) four-body repulsion
\beq{EqFourBody}
\scH\sub{4b} = V_4 \sum_{\!\!\!ijkl \in \square\!\!\!} (n_i - \tfrac{1}{2})(n_j - \tfrac{1}{2})(n_k - \tfrac{1}{2})(n_l - \tfrac{1}{2})\punc{,}
\eeq
where the sum is over sites \(ijkl\) around a square plaquette.

In this case, there are two continuous transitions, with density changing from $\langle n_i \rangle = \frac{1}{2}$ to $0 < \lvert\langle n_i \rangle - \frac{1}{2} \rvert < \frac{1}{2}$ and then to \(\lvert\langle n_i \rangle - \frac{1}{2} \rvert = \frac{1}{2}\), as \(V/t\) is increased. (The nature of the transitions is clear even for the small system sizes accessible in ED, because the order parameter commutes with the Hamiltonian.) The transition into the vacuum or fully-occupied (vacuum of holes) state is described by the standard critical theory for the vacuum transition of bosons,\cite{SachdevBook} while the transition at lower \(V/t\) involves spontaneous breaking of particle--hole symmetry within the superfluid, and is described by the critical theory of \refsec{SecCriticalTheory}. In cases where the total particle number is fixed, the latter transition would lead to phase separation into regions with differing densities.

The additional interaction \(\scH\sub{4s}\) in the classical spin model, defined in \refeq{EqFourSpin1}, may be viewed as the equivalent of \(\scH\sub{4b}\). To see this, recall that strings of flipped spins are equivalent to bosons, and that these occur at low density near the transition to saturation (bosonic vacuum). Two strings passing through a tetrahedron \(t\) change its total spin from \(+\frac{4}{\sqrt{3}}\hat{\boldsymbol{z}}\) to \(-\frac{4}{\sqrt{3}}\hat{\boldsymbol{z}}\), and so the interaction \(\scH\sub{4s}\) (with \(V_4 > 0\)) amounts to an energy penalty when four strings are in close proximity (passing through two tetrahedra on opposite sides of the same hexagon).


\begin{thebibliography}{99}
\bibitem{FootnoteOtherInteractions}
This remark notwithstanding, we note that some other apparently four-spin interactions reduce to a sum of two-spin interactions under projection to the ice-rule states. For example, an interaction similar to \(\scH\sub{4s}\) but giving \(\pm V_4\) when the two tetrahedra are polarized in the same (\(+\)) or opposite (\(-\)) directions along \(\hat{\boldsymbol{z}}\) can be written as \(S_t^z S_{t'}^z\) and is therefore a sum of two-spin interactions.
\bibitem{FootnoteBragg} {There are Bragg peaks at wavevector \(hh\ell\) (in units corresponding to the conventional cubic unit cell), for $h$, $\ell$ integers and $h+\ell$ even, except those where $h=0$ or where $h$ and $\ell$ are even but $h+\ell$ is not divisible by $4$.}
\end{thebibliography}
\end{document}